\documentclass[12pt,tightenlines,eqsecnum,floats,showpacs,nofootinbib,amsmath,amssymb,aps,prd]{revtex4}

\usepackage{graphicx}
\usepackage{amsmath,amssymb}
\usepackage{subfigure}

\def\be{\begin{equation}}
\def\ee{\end{equation}}
\def\ba{\begin{eqnarray}}
\def\ea{\end{eqnarray}}

\def\f{\frac}
\def\sgn{\mathrm{sgn}}
\def\l{\lambda}
\def\g{{\rm grav}}
\def\m{{\rm matt}}

\def\H{\mathcal{H}}
\def\Hp{\mathcal{H}_{\rm phy}}

\def\Hk{\H_{\rm kin}}
\def\Hkg{\Hk^{\rm grav}}
\def\V{\mathcal{V}}
\def\gen{{\rm Gen}}
\def\res{{\rm Res}}
\def\P{\hat{\mathbb{P}}}
\def\WDW{\rm WDW\,\,}
\def\ul{\underline}

\def\q{{}^o\!q}
\def\e{{}^o\!e}
\def\w{{}^o\!\omega}
\def\pT{p_{(T)}}
\def\T{T}

\def\dd{{\rm d}}
\def\R{\mathbb{R}}
\def\r{\rho}
\def\rp{\rho_{\rm Pl}}
\def\rcr{\rho_{\rm crit}}
\def\l{\lambda}
\def\lp{\ell_{\rm Pl}}

\begin{document}

%\preprint{\vbox{\baselineskip=12pt \rightline{IGPG-06/06-}
%\rightline{gr-qc/0607039}
%}}

\title{Loop quantum cosmology of Bianchi I models}
%\rapid{Bianchi I space-times in loop quantum cosmology}

\author{Abhay Ashtekar}
\email{ashtekar@gravity.psu.edu}
\author{Edward Wilson-Ewing}
\email{wilsonewing@gravity.psu.edu} \affiliation{Institute for
Gravitation and the Cosmos, \& Physics Department, The
Pennsylvania State University, University Park, PA 16802, U.S.A.}

\begin{abstract}

The ``improved dynamics'' of loop quantum cosmology is extended to
include anisotropies of the Bianchi I model. As in the isotropic
case, a massless scalar field serves as a relational time
parameter. However, the extension is non-trivial because one has
to face several conceptual subtleties as well as technical
difficulties. These include: a better understanding of the
relation between loop quantum gravity (LQG) and loop quantum
cosmology (LQC); handling novel features associated with the
non-local field strength operator in presence of anisotropies; and
finding dynamical variables that make the action of the
Hamiltonian constraint manageable. Our analysis provides a
conceptually complete description that overcomes limitations of
earlier works. We again find that the big bang singularity is
resolved by quantum geometry effects but, because of the presence
of Weyl curvature, Planck scale physics is now much richer than in
the isotropic case. Since the Bianchi I models play a key role in
the Belinskii, Khalatnikov, Lifshitz (BKL) conjecture on the
nature of generic space-like singularities in general relativity,
the quantum dynamics of Bianchi I cosmologies is likely to provide
considerable intuition about the fate of generic space-like
singularities in quantum gravity. Finally, we show that the
quantum dynamics of Bianchi I cosmologies projects down
\emph{exactly} to that of the Friedmann model. This opens a new
avenue to relate more complicated models to simpler ones, thereby
providing a new tool to relate the quantum dynamics of LQG to that
of LQC.

\end{abstract}

\pacs{98.80Qc,04.60.Pp, 04.60.-m}
% Quantum Cosmology; Loop Quantum Gravity, Quantum Gravity

\maketitle

\section{Introduction}
\label{s1}

Loop quantum gravity (LQG) \cite{alrev,crbook,ttbook} is a
non-perturbative, background independent approach to the
unification of general relativity and quantum physics. One of its
key features is that space-time geometry is treated quantum
mechanically from the beginning. Loop quantum cosmology (LQC)
\cite{mbrev,aa-rev} is constructed by applying methods of LQG to
mini-superspaces obtained by a symmetry reduction of general
relativity. In the homogeneous, isotropic cosmological models with
a massless scalar field, quantum geometry effects of LQG have been
shown to create a new repulsive force in the Planck regime. The
force is so strong that the big bang is replaced by a specific
type of quantum bounce. The force rises very quickly once the
scalar curvature reaches $\sim - 0.15\pi/\lp^2$ (or matter density
$\r$ reaches $\sim 0.01\,\rp$) to cause the bounce but also dies
very quickly after the bounce once the scalar curvature and the
density fall below these values. Therefore outside the Planck
regime the quantum space-time of LQC is very well approximated by
the space-time continuum of general relativity. This scenario is
borne out in the k=0, $\Lambda$=0 models,
\cite{mb1,abl,aps1,aps2,aps3,acs,cs,kl}, $\Lambda\not=$0 models
\cite{bp,ap}, the k=$1$ closed model \cite{apsv,warsaw1}, k=$ -1$
open model \cite{kv2} and the k=0 model with an inflationary
potential with phenomenologically viable parameters \cite{aps4}.
Going beyond the big-bang and big-crunch singularities, LQC has
also been used to argue that its quantum geometry effects resolve
\emph{all} strong curvature singularities in homogeneous,
isotropic situations in which matter is a perfect fluid with an
equation of state of the standard type, $p= p(\rho)$ \cite{ps}.
(For recent reviews, see, e.g., \cite{aa-badhonef,aa-saloniki}.)
Finally, recent investigations \cite{hybrid1,hybrid2} of Gowdy
models, which have an infinite number of degrees of freedom, also
indicate that the big-bang is replaced by a quantum bounce.

Detailed and viable quantum theories were constructed in the
homogeneous, isotropic case using the so-called ``$\bar\mu$''
scheme. A key open question has been whether or not the
qualitative features of their Planck scale physics will persist in
more realistic situations in which these strong symmetry
assumptions do not hold exactly. A first step in this direction is
to retain homogeneity and extend the ``improved dynamics'' of
\cite{aps3} to anisotropic situations. In the isotropic case,
there is only one non-trivial curvature invariant, the
(space-time) scalar curvature (or, equivalently, matter density).
In anisotropic situations Weyl curvature is non-zero and it too
diverges at the big bang. Therefore, now one can enter the Planck
regime in several inequivalent ways which suggests that the Planck
scale physics would now be much richer.

In this paper we will continue the LQC explorations of this issue by
analyzing in detail the simplest of anisotropic models, the Bianchi
I cosmologies. (Previous work on this model is discussed below.) As
in the isotropic case we will use a massless scalar field as the
matter source, and it will continue to provide the ``relational'' or
``internal'' time a la Leibniz with respect to which other physical
quantities of interest ---e.g., curvatures, shears, expansion and
matter density--- ``evolve''. Again, as in the isotropic case, the
framework can be further extended to accommodate additional matter
fields in a rather straightforward fashion.

Although the Bianchi I models are the simplest among anisotropic
cosmologies, results obtained in the context of the Belinskii,
Khalatnikov, Lifshitz (BKL) conjecture \cite{bkl1,bkl2} suggest
that they are perhaps the most interesting ones for the issue of
singularity resolution. The BKL conjecture states that, as one
approaches space-like singularities in general relativity, terms
with time derivatives would dominate over those with spatial
derivatives, implying that the asymptotic dynamics would be well
described by an ordinary differential equation. By now
considerable evidence has accumulated in favor of this conjecture
\cite{bmbkl,bgimw,aelu,dgbkl,ar}. For the case when the matter
source is a massless scalar field in full general relativity
without any symmetry assumption, these results suggest that, as
the system enters the Planck regime, dynamics along any fixed
spatial point would be well described by a Bianchi I metric.
Therefore understanding the fate of Bianchi I models in LQC could
provide substantial intuition for what happens to generic
space-like singularities in LQG \cite{dg,ahs}.

Indeed, in cosmological contexts where one has approximate
homogeneity, a natural strategy in full LQG is to divide the
spatial 3-manifold into small, elementary cells and assume that
there is homogeneity in each cell, with fields changing slowly as
one moves from one cell to the next. (For an exploration along
these lines in the older ``$\mu_o$ scheme,'' see \cite{mb-grg}.)
Now, if one were to assume that geometry in each elementary cell
is also isotropic, then the Weyl tensor in each cell ---and
therefore everywhere--- would be forced to be zero. A natural
strategy to accommodate realistic, non-vanishing Weyl curvature
would be to use Bianchi I geometry in each cell and let the
parameters $k_i$ vary slowly from one cell to another. In this
manner, LQC of the Bianchi I model can pave way to the analysis of
the fate of generic space-like singularities of general relativity
in full LQG.

Because of these potential applications, Bianchi I models have
already drawn considerable attention in LQC (see in particular
\cite{mbBI,chiou,cv,mb-stable,mbmmp,szulc}). During these
investigations, groundwork was laid down which we will use
extensively. However, in the spatially non-compact context (i.e.,
when the spatial topology is $\R^3$ rather than $\mathbb{T}^3$), the
construction of the quantum Hamiltonian constraint turned out to be
problematic. The Hamiltonian constraint used in the early work has
the same difficulties as those encountered in the ``$\mu_o$-scheme''
in the isotropic case (see, e.g., \cite{cs}, or Appendix B of
\cite{aa-badhonef}). More recent papers have tried to overcome these
limitations by mimicking the ``$\bar\mu$'' scheme used successfully
in the isotropic case. However, to make concrete progress, at a key
point in the analysis a simplifying assumption was made without a
systematic justification.%
\footnote{In the isotropic case, ``improved'' dynamics \cite{aps3}
required that $\bar\mu$ be proportional to $1/\sqrt{|p|}$. In the
anisotropic case, one has three $p_i$ and quantum dynamics
requires the introduction of three $\bar\mu_i$. In the Bianchi I
case now under consideration, it was simply assumed
\cite{chiou,cv,szulc} that $\bar\mu_i$ be proportional to
$1/\sqrt{|p_i|}$. We will see in section \ref{s3.2} that a more
systematic procedure leads to the conclusion that the correct
generalization of the isotropic result is more subtle. For
example, $\bar\mu_1$ is proportional to $\sqrt{|p_1|/|p_2p_3|}$.}
Unfortunately, it leads to quantum dynamics which depends, even to
leading order, on the choice of an auxiliary structure (i.e., the
fiducial cell) used in the construction of the Hamiltonian framework
\cite{szulc}. This is a major conceptual drawback. Also, the final
results inherit certain features that are not physically viable
(e.g. the dependence of the quantum bounce on ``directional
densities'' in \cite{chiou, cv}). We will provide a systematic
treatment of quantum dynamics that is free from these drawbacks.

To achieve this goal one has to overcome rather non-trivial
obstacles which had stalled progress for the past two years. This
requires significant new inputs. The first is conceptual: we will
sharpen the correspondence between LQG and LQC that underlies the
definition of the curvature operator $\hat{F}_{ab}^i$ in terms of
holonomies. The holonomies we are led to use in this construction
will have a non-trivial dependence on triads, stemming from the
choice of loops on which they are evaluated (see footnote 1). As a
result, at first it seems very difficult to define the action of the
resulting quantum holonomy operators. Indeed this was the primary
technical obstacle that forced earlier investigations to take
certain short cuts ---the assumption mentioned above--- while
defining $\hat{F}_{ab}^i$. The second new input is the definition of
these holonomy operators without having to take a recourse to such
short cuts. But then the resulting Hamiltonian constraint appears
unwieldy at first. The third major input is a rearrangement of
configuration variables that makes the constraint tractable both
analytically, as in this paper, and for the numerical work in
progress \cite{ht}.

Finally, we will find that the resulting Hamiltonian constraint has
a striking feature which could provide a powerful new tool in
relating the quantum dynamics of more complicated models to that of
simpler models. It turns out that, in LQC, there is a well-defined
projection from the Bianchi I physical states to the Friedmann
physical states which maps the Bianchi I quantum dynamics
\emph{exactly} to the isotropic quantum dynamics. Previous
investigations of the relation between quantum dynamics of a more
complicated model to that of a simpler model generally began with an
embedding of the Hilbert space $\H_{\rm Res}$ of the more restricted
model in the Hilbert space $\H_{\rm Gen}$ of the more general model
(see, e.g., \cite{kr,rt}). In generic situations, the image of
$\H_{\rm Res}$ under this embedding was not left invariant by the
more general dynamics on $\H_{\rm Gen}$. This led to a concern that
the physics resulting from first reducing and then quantizing may be
completely different from that obtained by quantizing the larger
system and regarding the smaller system as its sub-system. The new
idea of projecting from $\H_{\rm Gen}$ to $\H_{\rm Res}$ corresponds
to ``integrating out the degrees of freedom that are inaccessible to
the restricted model'' while the embedding $\H_{\rm Res}$ in to
$\H_{\rm Gen}$ corresponds to ``freezing by hand'' these extra
degrees of freedom. Classically, both are equally good procedures
and in fact the embedding is generally easier to construct. However,
in quantum mechanics it is more appropriate to integrate out the
``extra'' degrees of freedom. In the present case, one ``integrates
out'' anisotropies to go from the LQC of the Bianchi I models to
that of the Friedmann model. This idea was already proposed and used
in \cite{bhm} in a perturbative treatment of anisotropies in locally
rotationally symmetric, diagonal, Bianchi I model. We extend that
work in that we consider the full quantum dynamics of diagonal
Bianchi I model without additional symmetries and, furthermore, use
the analog of the ``$\bar\mu$ scheme'' in which the quantum
constraint is considerably more involved than in the
``$\mu_o$-type'' scheme used in \cite{bhm}. The fact that the LQC
dynamics of the Friedmann model is recovered exactly provides some
concrete support for the hope that LQC may capture the essential
features of full LQG, as far as the quantum dynamics of the
homogeneous, isotropic degree of freedom is concerned.

The material is organized as follows. We will begin in section
\ref{s2} with an outline of the classical dynamics of Bianchi I
models. This overview will not be comprehensive as our goal is only
to set the stage for the quantum theory which is developed in
section \ref{s3}. In section \ref{s4} we discuss three key
properties of quantum dynamics: the projection map mentioned above,
agreement of the LQC dynamics with that of the Wheeler DeWitt theory
away from the Planck regime and effective equations. (The isotropic
analogs of these equations approximate the full LQC dynamics of
Friedmann models extremely well.) In section \ref{s4} we summarize
the main results and discuss some of their ramifications. The
Appendix \ref{a1} discusses parity type discrete symmetries which
play an important role in the analysis of quantum dynamics.

\section{Hamiltonian Framework}
\label{s2}

In this section we will summarize those aspects of the classical
theory that will be needed for quantization. For a more complete
description of the classical dynamics see, e.g., \cite{rs-book,
mbBI,chiou,cv}. %In the first subsection we will review the dynamics of this
%model and, in the second, we will examine how the Hamiltonian
%constraint responds to changes of orientation of triads.

%\subsection{Classical Dynamics}
%\label{s2.1}

Our space-time manifold $M$ will be topologically $\R^4$. As is
standard in the literature on Bianchi models, we will restrict
ourselves to \emph{diagonal} Bianchi I metrics. Then one can fix
Cartesian coordinates $\tau, x_i$ on $M$ and express the space-time
metric as:
\be \label{metric} ds^2 = -N^2 d\tau^2 + a_1^2 \: dx_1^2 + a_2^2 \:
dx_2^2 + a_3^2 \: dx_3^2\, , \ee
where $N$ is the lapse and $a_i$ are the directional scale factors.
Thus, the dynamical degrees of freedom  are encoded in three
functions $a_i(\tau)$ of time. Bianchi I symmetries permit us to
rescale the three spatial coordinates $x_i$ by independent
constants. Under $x_i \rightarrow \alpha_i x_i$, the directional
scale factors transform as%
\footnote{Here and in what follows there is no summation over
repeated indices if they are all contravariant or all covariant. On
the other hand, a covariant index which is contracted with a
contravariant one is summed over 1,2,3.}
$a_i\rightarrow\alpha_i^{-1} a_i$. Thus, the numerical value of a
directional scale factor, say $a_1$, is not an observable; only
ratios such as $a_1(\tau)/a_1(\tau')$ are. The matter source will
be a massless scalar field which will serve as the relational or
internal time. Therefore, it is convenient to work with a harmonic
time function, i.e. to ask that $\tau$ satisfy $\Box \tau =0$.
From now on we will work with this choice.

Since the spatial manifold is non-compact and all fields are
spatially homogeneous, to construct a Lagrangian or a Hamiltonian
framework one has to introduce an elementary cell $\V$ and
restrict all integrations to it \cite{abl}. We will choose $\V$ so
that its edges lie along the fixed coordinate axis $x_i$. As in
the isotropic case, it is also convenient to fix a fiducial flat
metric $\q_{ab}$ with line element
\be ds_o^2 = dx_1^2 + dx_2^2 + dx_3^2\, . \ee
We will denote by $\q$ the determinant of this metric, by $L_i$ the
lengths of the three edges of $\V$ as measured by $\q_{ab}$, and by
$V_o = L_1L_2L_3$ the volume of the elementary cell $\V$ also
measured using $\q_{ab}$. Finally, we introduce fiducial co-triads
$\w_a^i = D_a x^i$ and the triads $\e^a_i$ dual to them. Clearly
they are adapted to the edges of $\V$ and are compatible with
$\q_{ab}$ (i.e., satisfy $\q_{ab} = \w_a^i\w_b^j \delta_{ij}$). As
noted above, Bianchi I symmetries allow each of the three
coordinates to be rescaled by an independent constant $\alpha_i$.
Under these rescalings, $x_i \rightarrow x^\prime_i = \alpha_i x_i$,
co-triads transform as $\w_a^{\prime\, i} = \alpha_i \w_a^i$, and
triads $\e^a_i$ are rescaled by inverse powers of $\alpha_i$. The
fiducial metric is transformed to $\q^\prime_{ab}$ defined by
$ds_o^{\prime\,2} := \alpha_1^2 dx_1^2 +\alpha_2^2 dx_2^2+\alpha_3^2
dx_3^2$. \emph{We must ensure that our physical results do not
change under these rescalings.} Finally, the physical co-triads are
given by $\omega_a^i = a^i \w_a^i$ and the physical 3-metric
$q_{ab}$ is given by $q_{ab} = \omega_a^i\omega_b^j\, \delta_{ij}$.

With these fiducial structures at hand, we can now introduce the
phase space. Recall first that in LQG the canonical pair consists of
an ${\rm SU(2)}$ connection $A_a^i$ and a triad $E^a_i$ of density
weight one. Using the Bianchi I symmetry, from each gauge
equivalence class of these pairs we can select one and only one,
given by:
\be \label{var} A_a^i =: c^i (L^i)^{-1}\: \w_a^i, \qquad
\mathrm{and} \qquad E^a_i \,\equiv\, \sqrt q\, e^a_i\, =:\, p_i
L_i V_o^{-1} \sqrt{\q}\: \e^a_i, \ee
where $c_i, p_i$ are constants and $q= (p_1p_2p_3)\, \q\,
V^{-1}_o$ is the determinant of the physical spatial metric
$q_{ab}$. \emph{Thus the connections $A_a^i$ are now labelled by
three parameters $c^i$ and the triads $E^a_i$ by three parameters
$p_i$.} If $p_i$ are positive, the physical triad $e^a_i$ and the
fiducial triad $\e^a_i$ have the same orientation. A change in
sign of, say, $p_1$ corresponds to a change in the orientation of
the physical triad brought about by the flip $e_1^a \rightarrow
-e_1^a$. These flips are gauge transformations because they do not
change the physical metric $q_{ab}$. The momenta $p_i$ are
directly related to the directional scale factors:
\be \label{a&p} p_1 = \sgn(a_1)|a_2a_3|\, L_2L_3, \qquad p_2 =
 \sgn(a_2)|a_1a_3|\,L_1L_3, \qquad p_3= \sgn(a_3)|a_1a_2|\, L_1L_2\, .
\ee
where we take the directional scale factor $a_i$ to be positive if
the triad vector $e^a_i$ is parallel to $\e^a_i$ and negative if it
is anti-parallel. As we will see below, in any solution to the field
equations, the connection components $c_i$ are directly related to
the time derivatives of $a_i$.

The factors of $L_i$ in (\ref{var}) ensure that this parametrization
is unchanged if the fiducial co-triad, triad and metric are rescaled
via $x_i \rightarrow \alpha_i\,x_i$. However, the parametrization
does depend on the choice of the cell $\V$. Thus the situation is
the same as in the isotropic case \cite{abl}. (The physical fields
$A_a^i$ and $E^a_i$ are of course insensitive to changes in the
fiducial metric \emph{or} the cell.) To evaluate the symplectic
structure of the symmetry reduced theory, as in the isotropic case
\cite{abl}, we begin with the expression of the symplectic structure
in the full theory and simply restrict the integration to the cell
$\V$. The resulting (non-vanishing) Poisson brackets are given by:
\be \label{pb} \{c^i, p_j\} = 8\pi G\gamma\,\delta^i_j\, . \ee

To summarize, the phase space in the Bianchi I model is six
dimensional, coordinatized by pairs $c^i, p_i$, subject to the
Poisson bracket relations (\ref{pb}). This description is tied to
the choice of the fiducial cell $\V$ but is insensitive to the
choice of fiducial triads, co-triads and metrics.

Next, let us consider constraints. The full theory has a set of
three constraints: the Gauss, the diffeomorphism and the
Hamiltonian constraints. It is straightforward to check that,
because we have restricted ourselves to diagonal metrics and fixed
the internal gauge, the Gauss and the diffeomorphism constraints
are identically satisfied. We are thus left with just the
Hamiltonian constraint. Its expression is obtained by restricting
the integration in the full theory to the fiducial cell $\V$:
\be \label{ham1} \mathcal{C}_H = \mathcal{C}_\g + \mathcal{C}_\m =
\int_\mathcal{V} N\,\left( \mathcal{H}_\g+\mathcal{H}_\m\right)\,
\dd^3x \ee
where $N$ is the lapse function and the gravitational and the
matter parts of the constraint densities are given by
%
%\footnote{We have dropped a term from $\mathcal{H}_g$ that contains the Gauss
%constraint which has already been gauge-fixed and is therefore 0.}
%
\be \mathcal{H}_\g = \f{E^a_iE^b_j}{16\pi
G\sqrt{|q|}}\left(\epsilon^{ij}{}_k F_{ab}{}^k
-2(1+\gamma^2)K_{[a}^iK_{b]}^j\right) \qquad \mathrm{and} \qquad
\mathcal{H}_\m = \sqrt{q} \: \rho_\m. \ee
Here $\gamma$ is the Barbero-Immirzi parameter, %which is fixed to
%0.2375 due to black hole entropy calculations \cite{bh-entropy}
$F_{ab}{}^k$ is the curvature of the connection $A_a^i$, given by
\be F_{ab}{}^k = 2\partial_{[a}A_{b]}{}^k +
\epsilon_{ij}{}^kA_a^iA_b^j\, ,\ee
$K_a^i$ is related to the extrinsic curvature $K_{ab}$ via $K_a^i
= K_{ab}e^{bi}$ and $\rho_\m$ is the energy density of the matter
fields. In general, $A_a^i$ is related to $K_a^i$ and the spin
connection $\Gamma_a^i$ defined by the triad $e^a_i$ via $A_a^i =
\Gamma_a^i + \gamma K_a^i$. However, because Bianchi I models are
spatially flat, $\Gamma_a^i =0$ in the gauge chosen in
(\ref{var}), whence $A_a^i = \gamma K_a^i$. This property and the
fact that spatial derivatives of $K_a^i$ vanish by the Bianchi I
symmetry leads us to the relation
\be 2K_{[a}^iK_{b]}^j=\gamma^{-2}\epsilon^{ij}{}_kF_{ab}{}^k\, . \ee
Therefore, the gravitational part of the Hamiltonian constraint can
be simplified:
\begin{align} \mathcal{H}_\g &= -\f{E^a_iE^b_j}{16\pi G\gamma^2\sqrt{q}}
\,\epsilon^{ij}{}_k\,F_{ab}{}^k \nonumber \\ & = -
\f{\sqrt{{}^oq}}{8\pi
G\gamma^2\sqrt{p_1p_2p_3}\,V_o}\,\,\left(p_1p_2c_1c_2
+p_1p_3c_1c_3+p_2p_3c_2c_3\right), \end{align}

Finally, recall that our matter field is a massless scalar field
$\T$. The matter energy density of the scalar field $T$ is given by
$\rho_\m = \pT^2 /2V^2$, where $V=\sqrt{|p_1p_2p_3|}$ is the
physical volume of the elementary cell. Our choice of harmonic time
$\tau$ implies that the lapse function is given by
$N=\sqrt{|p_1p_2p_3|}$. With these choices the constraint
(\ref{ham1}) simplifies further:
\begin{align} \label{ham2} \mathcal{C}_H &= \int_\mathcal{V}
\left(-\f{E^a_iE^b_jV_o}{16\pi G \gamma^2 \sqrt{{}^oq}}
\epsilon_{ij}{}^kF_{ab}^k+\f{\sqrt{{}^oq}}{V_o}
\f{p_\T^2}{2}\right)\, d^3x  \\
&  = -\f{1}{8\pi G\gamma^2}
\left(p_1p_2c_1c_2+p_1p_3c_1c_3+p_2p_3c_2c_3\right)
+\f{p_\T^2}{2}. \label{ham3}\end{align}
Physical states of the classical theory lie on the constraint
surface $\mathcal{C}_H =0$. The time evolution of each $p_i$ and
$c_i$ is obtained by taking their Poisson bracket with
$\mathcal{C}_H$.
\be \label{dp1} \f{\dd p_1}{\dd\tau} \,=\, \{p_1,\mathcal{C}_H\}\,
=\, -8\pi G\gamma \f{\partial\mathcal{C}_H}{\partial c_1}\, =\,
\f{p_1}{\gamma}\left(p_2c_2+p_3c_3 \right); \ee
\be \label{dc1} \f{\dd c_1}{\dd\tau}\, =\, \{c_1,\mathcal{C}_H\}\,
=\, 8\pi G \gamma \f{\partial \mathcal{C}_H}{\partial p_1}\, =\,
\f{-c_1}{\gamma} \left(p_2c_2+p_3c_3\right). \ee
The four other time derivatives can be obtained via permutations.
Although the phase space coordinates $c^i, p_i$ themselves depend on
the choice of the fiducial cell $\V$, the dynamical equations for
$A_a^i$ and $E^a_i$ ---and hence also for the physical metric
$q_{ab}$ and the extrinsic curvature $K_{ab}$--- that follow from
(\ref{dp1}) and (\ref{dc1}) \emph{are independent of this choice.}

Combining Eqs. (\ref{a&p}), (\ref{dp1}) and (\ref{dc1}), one finds
\be c_i = \gamma L_i\, V_o^{-1}\, (a_1a_2a_3)^{-1}\,\f{\dd
a_i}{\dd\tau} \ee
It is instructive to relate the $c_i$ to the directional Hubble
parameters $H_i = \dd \ln a_i/\dd t$ where \emph{$t$ is the proper
time}, corresponding to the lapse function $N_{(t)}=1$. Since $t$ is
related to the harmonic time $\tau$ via $N\dd\tau = N_{(t)}\dd t$
\be \label{tau} \f{\dd}{\dd t}= \f{1}{\sqrt{|p_1p_2p_3|}}\,\,
\f{\dd}{\dd\tau}. \ee
Therefore, we have
\be c_i \,=\, \gamma \, L_i \, \f{\dd a_i}{\dd t}\, =\, \gamma L_i
a_i H_i\,  \ee
where $L_ia_i$ is the length of the $i$th edge of $\V$ as measured
by the physical metric $q_{ab}$.

Next, it is convenient to introduce a mean scale factor
$a:=(a_1a_2a_3)^{1/3}$ which encodes the physical volume element but
ignores anisotropies. Then, the mean Hubble parameter is given by
\be \label{Hubble} H := \f{\dd \ln a}{\dd t} =
\f{1}{3}\,\left(H_1+H_2+H_3 \right), \qquad {\hbox \mathrm{where as
before}} \qquad H_i := \f{\dd \ln a_i}{\dd t}. \ee
Squaring Eq. (\ref{Hubble}) and using the implication
\be H_1H_2 + H_2H_3+H_3H_1 = 8\pi\, G \rho_\m \ee
of the Hamiltonian constraint, we obtain the generalized Friedmann
equation for Bianchi I space-times,
\be \label{friedC} H^2 = \f{8\pi G}{3}\rho_\m + \f{\Sigma^2}{a^6},
\ee
where
\be \label{Sigma} \Sigma^2 =
\f{a^6}{18}\left[(H_1-H_2)^2+(H_2-H_3)^2+(H_3-H_1)^2\right] \ee
is the shear term. The right hand side of (\ref{friedC}) brings out
the fact that the anisotropic shears $(H_i-H_j)$ contribute to the
energy density; they quantify the energy density in the
gravitational waves. Using the fact that our matter field has zero
anisotropic stress one can show that $\Sigma^2$ is a constant of the
motion \cite{cv}. If the space-time itself is isotropic, then
$\Sigma^2=0$ and Eq. (\ref{friedC}) reduces to the usual Friedmann
equation for the standard isotropic cosmology. These considerations
will be useful in interpreting quantum dynamics and exploring the
relation between the Bianchi I and Friedmann quantum Hamiltonian
constraints.

Next, let us consider the scalar field $\T$. Because there is no
potential for it, its canonically conjugate momentum $\pT$ is a
constant of motion (which, for definiteness, will be assumed to be
positive). Therefore, in any solution to the field equations $\T$
grows linearly in the harmonic time $\tau$. Thus, although $\T$ does
not have the physical dimensions of time, it is a good evolution
parameter in the classical theory. The form of the quantum
Hamiltonian constraint is such that $\T$ will also serve as a viable
internal time parameter in the quantum theory.

We will conclude with a discussion of discrete `reflection
symmetries' that will play an important role in the quantum theory.
(For further details see the Appendix.) In the isotropic case, there
is a single reflection symmetry, $\Pi(p)=-p$ which physically
corresponds to the orientation reversal $e^a_i \rightarrow -e^a_i$
of triads. These are large gauge transformations, under which the
metric $q_{ab}$ remains unchanged. The Hamiltonian constraint is
invariant under this reflection whence one can, if one so wishes,
restrict one's attention just to the sector $p\ge0$ of the phase
space. In the Bianchi I case, we have three reflections $\Pi_i$,
each corresponding to the flip of one of the triad vectors, leaving
the other two untouched (e.g., $\Pi_1 (p_1,p_2,p_3) = (-p_1,
p_2,p_3)$). As shown in \cite{bd}, the Hamiltonian flow is left
invariant under the action of each $\Pi_i$. Therefore, it suffices
to restrict one's attention to the positive octant in which all
three $p_i$ are non-negative: dynamics in any of the other seven
octants can be easily recovered from that in the positive octant by
the action of the discrete symmetries $\Pi_i$.

\begin{itemize}

\item \emph{Remark:} In the LQC literature on Bianchi I models,
    a physical distinction has occasionally been  made between
    the fiducial cells $\V$ which are ``cubical'' with respect
    to the fiducial metric $\q_{ab}$ and those that are
    ``rectangular.'' (In the former case all $L_i$ are equal.)
    However, given \emph{any} cell $\V$ one can always find a
    flat metric in our collection (\ref{metric}) with respect to
    which that $\V$ cubical. Using it as $\q_{ab}$ one would be
    led to call it cubical. Therefore the distinction is
    unphysical and the hope that the restriction to cubical
    cells may resolve some of the physical problems faced in
    \cite{chiou,cv} was misplaced.

\end{itemize}

\section{Quantum Theory}
\label{s3}

This section is divided into four parts. In the first, we briefly
recall quantum kinematics, emphasizing issues that have not been
discussed in the literature. In the second, we spell out a simple
but well-motivated correspondence between the LQG and LQC quantum
states that plays an important role in the definition of the
curvature operator $\hat{F}_{ab}{}^k$ in terms of holonomies.
However, the paths along which holonomies are evaluated depend in a
rather complicated way on the triad (or momentum) operators, whence
at first it seems very difficult to define these holonomy operators.
In the third subsection we show that geometric considerations
provide a natural avenue to overcome these apparent obstacles. The
resulting Hamiltonian constraint is, however, rather unwieldy to
work with. In the last sub-section we make a convenient redefinition
of configuration variables to simplify its action. The
simplification, in turn, will provide the precise sense in which the
singularity is resolved in the quantum theory.

\subsection{LQC Kinematics}
\label{s3.1}

We will summarize quantum kinematics only briefly; for details, see
e.g. \cite{chiou, cv}. Let us begin by specifying the elementary
functions on the classical phase space which are to have unambiguous
analogs in the quantum theory. In LQC this choice is directly
motivated by the structure of full LQG \cite{alrev,crbook,ttbook}.
As one might expect from the isotropic case \cite{abl,aps2}, the
elementary variables are the three momenta $p_i$ and holonomies
$h_i^{(\ell)}$ along edges parallel to the three axis $x_i$, where
$\ell L_i$ is the length of the edge with
respect to the fiducial metric $\q_{ab}$.%
\footnote{More precisely, the dimensionless number $\ell$ is the
length of the edge along which the holonomy is evaluated, measured
in the units of the length of the edge of $\V$ parallel to it. Since
$\ell$ is a ratio of lengths, its value does not depend on the
fiducial or any other metric.}
These functions are (over)complete in the sense that they suffice to
separate points of the phase space. Taking the $x_1$ axis for
concreteness, the holonomy $h_1^{(\ell)}$ has the form
\be \label{hol} h_1^{(\ell)} (c_1,c_2,c_3) = \cos \f{c_1\ell}{2}
\mathbb{I} + 2 \sin \f{c_1 \ell}{2} \tau_1 \ee
where $\mathbb{I}$ is the unit $ 2 \times 2$ matrix and $\tau_i$
constitute a basis of the Lie algebra of ${\rm SU(2)}$, satisfying
$\tau^i\tau^j= \textstyle{1\over 2} \epsilon^{ij}{}_k \tau^k -
\textstyle{1\over 4} \delta^{ij}\mathbb{I}$. Thus, the holonomies
are completely determined by almost periodic functions $\exp (i\ell
c_j)$ of the connection; they are called ``almost'' periodic because
$\ell$ is any real number rather than an integer. In quantum theory,
then, elementary operators $\hat{h}_i^{(\ell)}$ and $\hat{p}_i$ are
well-defined  and our task is to express other operators of physical
interest in terms of these elementary ones.

Recall that in the isotropic case it is simplest to specify the
gravitational sector of the kinematic Hilbert space in the triad of
$p$ representation: it consists of wave functions $\Psi(p)$ which
are symmetric under $p \rightarrow -p$ and have a finite norm:
$||\Psi||^2 = \sum_p |\Psi(p)|^2 <\infty $. In the Bianchi I case it
is again simplest to describe $\Hkg$ in the momentum representation.
Consider first a \emph{countable} linear combination,
\be \label{norm} |\Psi \rangle = \sum_{p_1,p_2,p_3} \Psi(p_1, p_2,
p_3)\, |p_1, p_2, p_3\rangle \quad {\rm with}\quad
\sum_{p_1,p_2,p_3}\, |\Psi (p_1,p_2,p_3)|^2 \, < \infty , \ee
of orthonormal basis states $|p_1, p_2, p_3\rangle$, where
\be \langle p_1, p_2, p_3\, |\, p_1', p_2', p_3'\rangle =
\delta_{p_1\, p_1'}\, \delta_{p_2\, p_2'} \, \delta_{p_3\, p_3'}\,
.\ee
Next, recall that on the classical phase space the three
reflections $\Pi_i$ represent large gauge transformations under
which physics does not change. They have a natural induced action
$\hat{\Pi}_i$ on the space of wave functions $\Psi(p_1,p_2,p_3)$.
(Thus, for example, $\hat\Pi_1 \Psi(p_1,p_2,p_3) = \Psi (-p_1,
p_2,p_3)$.) Physical observables commute with $\hat\Pi_i$.
Therefore, as in gauge theories, each eigenspace of $\hat{\Pi}_i$
provides a physical sector of the theory. %Since the image of the
%positive octant $p_i \ge 0$
%under this group is the full 3-dimensional space $\R^3$ spanned by
%all $p_i$, states $\Psi(p_1,p_2,p_3)$ in any eigenspace are
%completely determined by their restriction to the positive octant.
Since $\hat\Pi_i^2 = \mathbb{I}$, eigenvalues of $\hat\Pi_i$ are
$\pm 1$. For definiteness, as in the isotropic case, we will
assume that the wave functions $\Psi(p_1,p_2,p_3)$ are symmetric
under $\hat\Pi_i$. Thus, the gravitational part $\Hkg$ of the
kinematical Hilbert space is spanned by wave functions
$\Psi(p_1,p_2,p_3)$ satisfying
\be \label{parity} \Psi(p_1,p_2,p_3) = \Psi (|p_1|, |p_2|, |p_3|)
\ee
which have finite norm (\ref{norm}).

The basis states $|p_1,p_2,p_3\rangle$ are eigenstates of quantum
geometry: In the state $|p_1, p_2, p_3\rangle$ the face $S_i$ of the
fiducial cell $\V$ orthogonal to the axis $x_i$ has area $|p_i|$.
Note that although $p_i \in \R$, the orthonormality holds via
Kronecker deltas rather than the usual Dirac distributions; this is
why the LQC quantum kinematics is inequivalent to that of the
Schr\"odinger theory used in Wheeler DeWitt cosmology. Finally the
action of the elementary operators is given by:
\be \hat{p}_1\,|p_1,p_2,p_3\rangle = p_1\,|p_1,p_2,p_3\rangle \quad
\mathrm{and} \quad \widehat{\exp {i\ell c_1}}|p_1,p_2,p_3\rangle =
|p_1-8\pi \gamma G\hbar\ell,\, p_2,p_3\rangle \ee
and similarly for $\hat{p_2},\, \widehat{\exp i\ell c_2},\,
\hat{p_3}$ and $\widehat{\exp i\ell c_3}$.

The full kinematical Hilbert space $\Hk$ will be the tensor
product, $\Hk = \Hkg \otimes \Hk^\m$ where, as in the isotropic
case, we will set $\Hk^\m = L^2(\R, \dd\T)$ for the Hilbert space
of the homogeneous scalar field $\T$. On $\Hk^\m$, the operator
$\hat{\T}$ will act by multiplication and $\hat{p}_{(\T)}:=
-i\hbar \dd/\dd\T$ will act by differentiation. Note that \emph{we
can also use a ``polymer Hilbert space'' for $\Hk^\m$ spanned by
almost periodic functions of $\T$.} The quantum Hamiltonian
constraint (\ref{qham4}) will remain unchanged and our
construction of the physical Hilbert space will go through as it
is \cite{jl}.

\subsection{The curvature operator $\hat{F}_{ab}{}^k$}
\label{s3.2}

To discuss quantum dynamics, we have to construct the quantum analog
of the Hamiltonian constraint. Since there is no operator
corresponding to the connection coefficients $c_i$ on $\Hkg$, we
cannot use (\ref{ham3}) directly. Rather, as in the isotropic case
\cite{aps3}, we will return to the expression (\ref{ham2}) involving
curvature $F_{ab}{}^k$. Our task then is to find the operator on
$\Hkg$ corresponding to $F_{ab}{}^k$. As is usual in LQG, the idea
is to first express the curvature in terms of our elementary
variables ---holonomies and triads--- and then replace them by their
direct quantum analogs. Recall first that, in the classical theory,
the $a$-$b$ component of $F_{ab}{}^k$ can be written in terms of
holonomies around a plaquette (i.e., a rectangular closed loop whose
edges are parallel to two of the axes $x_i$):
\be \label{F1} F_{ab}{}^k = 2\lim_{Ar_\Box\rightarrow 0}\,\,
\mathrm{Tr}\left(\f{h_{\Box_{ij}}- \mathbb{I}} {Ar_\Box}
\:\tau^k\right)\,{}^o\omega_a^i\:{}^o\omega_b^j, \ee
where $Ar_\Box$ is the area of the plaquette $\Box$ and the holonomy
$h_{\Box_{ij}}$ around the plaquette $\Box_{ij}$ is given by
\be \label{holonomy} h_{\Box_{ij}}= {h_j^{(\bar\mu_j)}}^{-1}\,
{h_i^{(\bar\mu_i)}}^{-1}\, h_j^{(\bar\mu_j)}\, h_i^{(\bar\mu_i)}\,
\ee
where $\bar\mu_j\, L_j$ is the length of the $j$th edge of the
plaquette, as measured by the fiducial metric $\q_{ab}$. (There is
no summation over $i,j$.) Because the $Ar_{\Box}$ is shrunk to zero,
the limit is not sensitive to the precise choice of the closed
plaquette $\Box$. Now, in LQG the connection operator does not
exist, whence if we regard the right side of (\ref{F1}) as an
operator, the limit fails to converge in $\Hkg$. The non-existence
of the connection operator is a direct consequence of the underlying
diffeomorphism invariance \cite{aa-je} and is intertwined with the
fact that the eigenvalues of geometric operators ---such as the area
operator $\hat{Ar}_\Box$ associated with the plaquette under
consideration--- are purely discrete. Therefore, in LQC the
viewpoint is that the non-existence of the limit $Ar_\Box
\rightarrow 0$ in quantum theory is not accidental: quantum geometry
is simply telling us that we should shrink the plaquette not till
the area it encloses goes to zero, but rather only to the minimum
non-zero eigenvalue $\Delta\,\lp^2$ of the area operator (where
$\Delta$ is a dimensionless number). The resulting quantum operator
$\hat{F}_{ab}{}^k$ then inherits Planck scale non-localities.

\begin{figure}[tb]
  \begin{center}
  \subfigure[]
      {\includegraphics[width=2.5in]{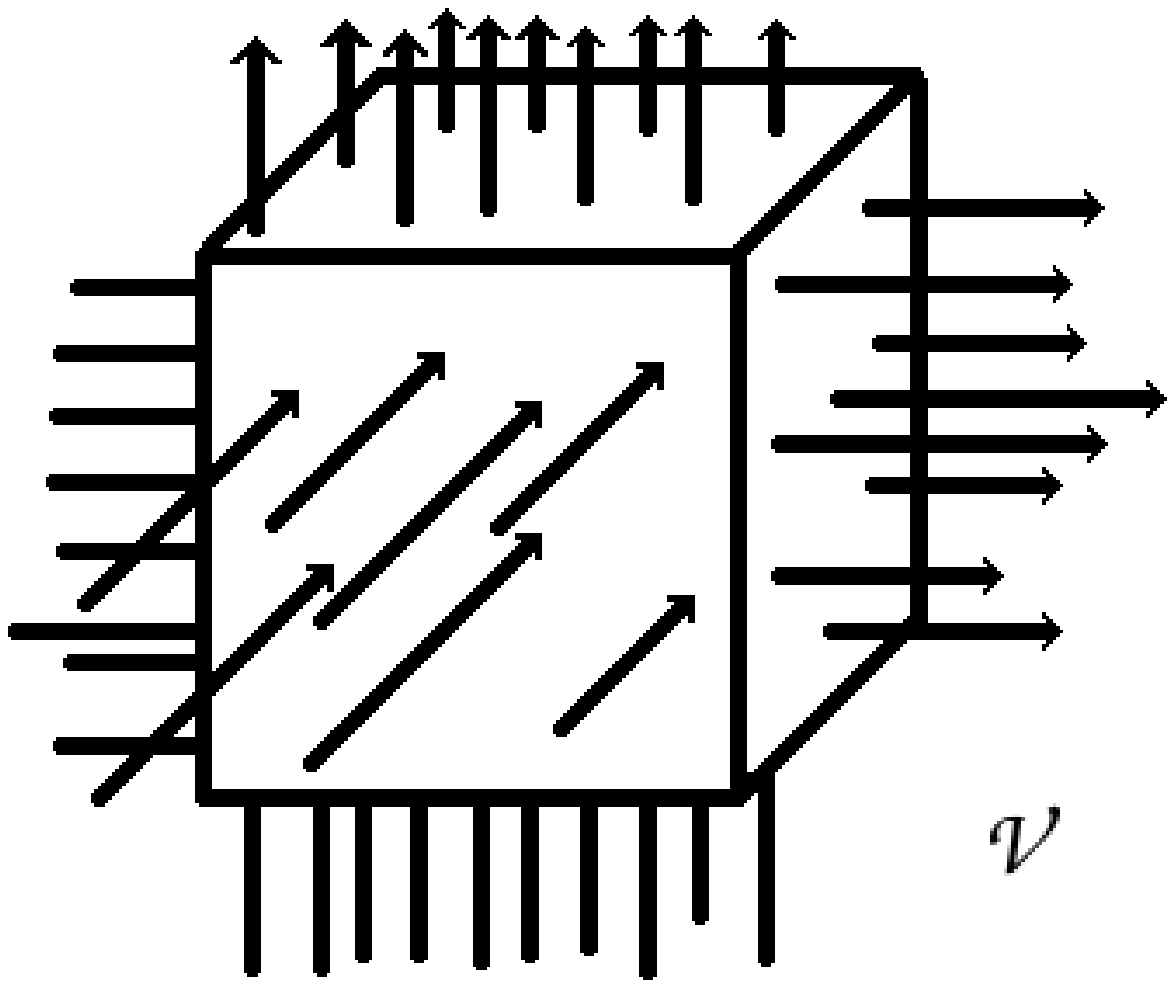}}
    \subfigure[]
      {\includegraphics[width=3.5in]{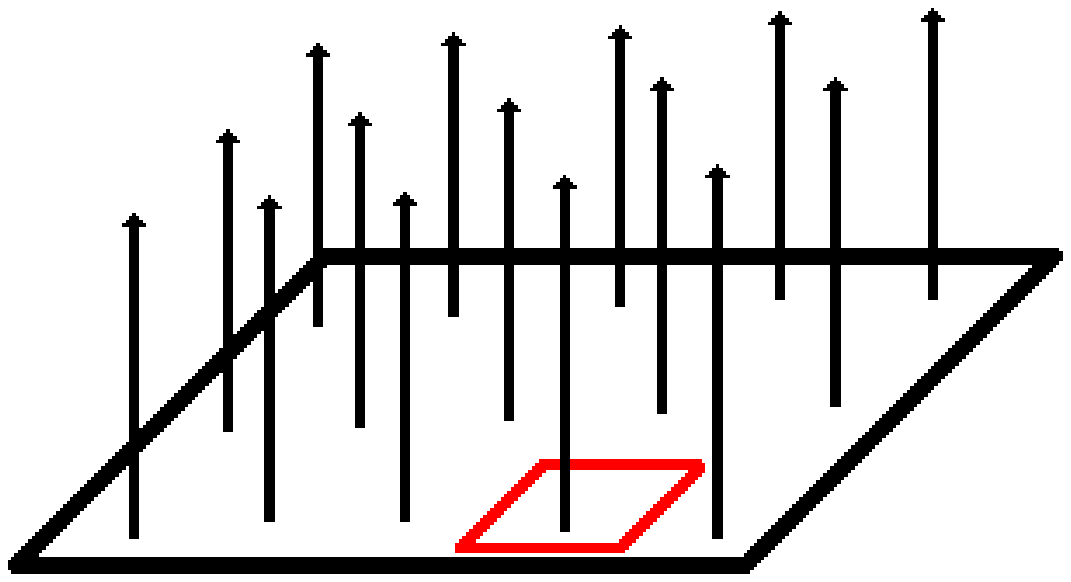}}
    \caption{Depiction of the LQG quantum geometry state corresponding to
    the LQC state $|p_1,p_2,p_3\rangle$. The LQG spin-network has edges parallel
    to the three axes selected by the diagonal Bianchi I symmetries, each carrying
    a spin label $j=1/2$.\,\,
    (a) Edges of the spin network traversing through the fiducial cell $\mathcal{V}$.
    (b) Edges of the spin network traversing the 1-2 face of $\mathcal{V}$ and
        an elementary plaquette associated with a single flux line. This plaquette
        encloses the smallest quantum, $\Delta\, \lp^2$, of area. The curvature
        operator $\hat{F}_{12}{}{}^k$ is defined by the holonomy around such a
        plaquette.}
  \label{fig2}
 \end{center}
\end{figure}

To implement this strategy in full LQG one must resolve a difficult
issue. If the plaquette is to be shrunk only to a finite size, the
operator on the right side of (\ref{F1}) would depend on what that
limiting plaquette is. So, which of the many plaquettes enclosing an
area $\Delta\,\lp^2$ should one use? Without a well-controlled gauge
fixing procedure, it would be very difficult to single out such
plaquettes, one for each 2-dimensional plane in the tangent space at
each spatial point. However, in the diagonal Bianchi I case now
under consideration, a natural gauge fixing is available and indeed
we have already carried it out. Thus, in the $i$-$j$ plane, it is
natural to choose a plaquette $\Box_{ij}$ so that its edges are
parallel to the $x_i$-$x_j$ axis. Furthermore, the underlying
homogeneity implies that it suffices to introduce the three
plaquettes at any one point in our spatial 3-manifold.

These considerations severely limit the choice of plaquettes
$\Box_{ij}$ but they do not determine the lengths of the two edges
in each of these plaquettes. To completely determine the
plaquettes, as in the isotropic case, we will use a simple but
well-motivated correspondence between kinematic states in LQG and
those in LQC. However, because of anisotropies, new complications
arise which require that the correspondence be made much more
precise. Fix a state $|p_1,p_2,p_3\rangle$ in $\Hkg$ of LQC. In
this state, the three faces of the fiducial cell $\V$ orthogonal
to the $x_i$-axis have areas $|p_i|$ in the LQC \emph{quantum}
geometry. This is the complete physical information in the ket
$|p_1,p_2,p_3\rangle$. How would this quantum geometry be
represented in full LQG? First, the macroscopic geometry must be
spatially homogeneous and we have singled out three axes with
respect to which our metrics are diagonal. Therefore,
semi-heuristic considerations suggest that the corresponding LQG
quantum geometry state should be represented by a spin network
consisting of edges parallel to the three axes (see Fig. 1(a)).
Microscopically this state is not exactly homogeneous. But the
\emph{coarse grained} geometry should be homogeneous. To achieve
the best possible coarse grained homogeneity, the edges should be
packed as tightly as is possible in the desired quantum geometry.
That is, each edge should carry the smallest non-zero label
possible, namely $j =1/2$.

For definiteness, let us consider the 1-2 face $S_{12}$ of the
fiducial cell $\V$ which is orthogonal to the $x_3$ axis (see Fig.
1(b)). Quantum geometry of LQG tells us that at each intersection
of any one of its edges with $S_{12}$, the spin network
contributes a quantum of area $\Delta\, \lp^2$ on this surface,
where $\Delta = 4\pi \gamma \sqrt{3}$ \cite{al5}. For this LQG
state to reproduce the LQC state $|p_1,p_2,p_3\rangle$ under
consideration $S_{12}$ must be pierced by $N_3$ edges of the LQG
spin network, where $N_3$ is given by
$$N_3\, \Delta\, \lp^2 = |p_3|\, .$$
Thus, we can divide $S_{12}$ into $N_3$ identical rectangles each
of which is pierced by exactly one edge of the LQG state, as in
Fig. 1(b). Any one of these elementary rectangles encloses an area
$\Delta \lp^2$ and provides us the required plaquette $\Box_{12}$.
Let the dimensionless lengths of the edges of these plaquettes  be
$\bar\mu_1$ and $\bar\mu_2$. Then their lengths with respect to
the fiducial metric $\q_{ab}$ are $\bar\mu_1L_1$ and
$\bar\mu_2L_2$. Since the area of $S_{12}$ with respect to
$\q_{ab}$ is $L_1L_2$, we have
$$ N_3\,\, \bar\mu_1 L_1\, \bar\mu_2L_2 = L_1L_2\, . $$
Equating the expressions of $N_3$ from the last two equations, we
obtain
\be \bar\mu_1 \bar\mu_2 = \f{\Delta\, \lp^2}{|p_3|}\, . \ee
This relation by itself does not fix $\bar\mu_1$ and $\bar\mu_2$.
However, repeating this procedure for the 2-3 face and the 3-1 face,
we obtain, in addition, two cyclic permutations of this last
equation and the three simultaneous equations do suffice to
determine $\bar\mu_i$:
\be \label{mubar} \bar\mu_1 = \sqrt\f{|p_1|\Delta\,\lp^2}{|p_2p_3|},
\quad \bar\mu_2 = \sqrt\f{|p_2| \Delta\,\lp^2}{|p_1p_3|}, \quad
\bar\mu_3 = \sqrt\f{|p_3|\Delta\,\lp^2}{|p_1p_2|}\, . \ee

To summarize, by exploiting the Bianchi I symmetries and using a
simple but well-motivated correspondence between LQG and LQC states
we have determined the required elementary plaquettes enclosing an
area $\Delta\,\lp^2$ on each of the three faces of the cell $\V$. On
the face $S_{ij}$, the plaquette is a rectangle whose sides are
parallel to the $x_i$ and $x_j$ axes and whose dimensionless lengths
are $\bar\mu_i$ and $\bar\mu_j$ respectively, given by
(\ref{mubar}). Note that (as in the isotropic case \cite{aps3}) the
$\bar\mu_i$ and hence the plaquettes are not fixed once and for all;
they depend on the LQC state $|p_1,p_2,p_3\rangle$ of quantum
geometry in a specific fashion. The functional form of this
dependence is crucial to ensure that the resulting quantum dynamics
is free from the difficulties encountered in earlier works.

Components of the curvature operator $\hat{F}_{ab}{}^k$ can now be
expressed in terms of holonomies around these plaquettes:
\be \label{F2} \hat{F}_{ab}{}^k = 2\,\,
\mathrm{Tr}\left(\f{h_{\Box_{ij}}- \mathbb{I}} {\Delta \lp^2}
\:\tau^k\right)\,{}^o\omega_a^i\:{}^o\omega_b^j, \ee
with
\be h_{\Box_{ij}} = {h_j^{(\bar\mu_j)}}^{-1}\,
{h_i^{(\bar\mu_i)}}^{-1}\, h_j^{(\bar\mu_j)}\, h_i^{(\bar\mu_i)}\,
, \ee
where $\bar\mu_j$ are given by (\ref{mubar}). (There is no summation
over $i,j$.) Using the expression (\ref{hol}) of holonomies, it is
straightforward to evaluate the right hand side. One finds:
\be \hat{F}_{ab}{}^k = \epsilon_{ij}{}^k\left(\f{\sin\bar\mu
c}{\bar\mu} \:{}^o\omega_a\right)^i\left(\f{\sin\bar\mu
c}{\bar\mu}\:{}^o\omega_b \right)^j, \ee
where the usual summation convention for repeated covariant and
contravariant indices applies and
\be \left(\f{\sin\bar\mu c}{\bar\mu}\: \w_a\right)^i=
\f{\sin\bar\mu^ic^i}{\bar\mu^i}\:\w_a^i, \ee
where there is now no sum over $i$. This is the curvature operator
we were seeking.\\

We will conclude with a discussion of the important features of this
procedure and of the resulting quantum dynamics.

1. In the isotropic case all $p_i$ are equal ($p_i = p$) whence our
expressions for $\bar\mu_i$ reduce to a single formula, $\bar\mu
=\sqrt{\Delta \lp^2/|p|}$. This is precisely the result that was
obtained in the ``improved dynamics'' scheme for the $k=0$ isotropic
models. Thus, we have obtained a generalization of that result to
Bianchi I models.

2. In both cases, the key observation is that the plaquette should
be shrunk till its area with respect to the physical ---rather than
the fiducial--- geometry is $\Delta\, \lp^2$. However, there are
also some differences. First, in the above analysis we set up and
used a correspondence between \emph{quantum} geometries of LQG and
LQC in the context of Bianchi I models. In contrast to the previous
treatment in the isotropic models \cite{aps3}, we did not have to
bring in classical geometry in the intermediate steps. In this
sense, even for the isotropic case, the current analysis is an
improvement over what is available in the literature.

3. A second difference between our present analysis and that of
\cite{aps3} is the following. Here, the semi-heuristic
representation of LQC states $|p_1,p_2,p_3 \rangle$ in terms of
spin networks of LQG suggested that we should consider spin
networks which pierce the faces of the fiducial cell $\V$ as in
Fig. 1(a). (As one would expect, these states are gauge
invariant.) The minimum non-zero eigenvalue of the area operator
on such states is $\Delta\, \lp^2$ with $\Delta =
4\sqrt{3}\pi\gamma$. This is \emph{twice} the absolute minimum of
non-zero eigenvalue on \emph{all} gauge invariant states. However,
that lower value is achieved on spin networks (whose edges are
again labelled by $j=1/2$ but) which do not pierce the surface but
rather intersect it from only one side. (In order for the state to
be gauge invariant, the edge then has to continue along a
direction tangential to the surface. For details, see \cite{al5}.)
Obvious considerations suggest that such states cannot feature in
homogeneous models. Since the discussion in the isotropic case
invoked a correspondence between LQG and LQC at a rougher level,
this point was not noticed and the value of $\Delta$ used in
\cite{aps3} was $2\sqrt{3}\pi \gamma$. We emphasize, however, that
although the current discussion is more refined, it is not a
self-contained derivation. A more complete analysis may well
change this numerical factor again.

4. On the other hand, we believe that the functional dependence of
$\bar\mu_i$ on $p_i$ is robust: As in the isotropic case this
dependence appears to be essential to make quantum dynamics
viable. Otherwise quantum dynamics can either depend on the choice
of the fiducial cell $\V$ even to leading order, or is physically
incorrect because it allows quantum effects to dominate in
otherwise ``tame'' situations, or both. The previous detailed,
quantum treatments of the Bianchi I model in LQC did not have this
functional dependence because they lacked the correspondence
between LQG and LQC we used. Rather, they proceeded by analogy. As
we noted above, in the isotropic case there is a single $\bar\mu$
and a single $p$ and the two are related by $\bar\mu =
\sqrt{\Delta \lp^2/|p|}$. The most straightforward generalization
of this relation to Bianchi I models is $\bar\mu_i = \sqrt{\Delta
\lp^2/|p_i|}$. This expression was simply postulated and then used
to construct quantum dynamics \cite{chiou, cv}. The resulting
analysis has provided a number of useful technical insights.
However, this quantum dynamics suffers from the problems mentioned
above \cite{szulc}. The possibility that the correct
generalization of the isotropic results to Bianchi I models may be
given by (\ref{mubar}) was noted in \cite{kv, mb-stable} and in
the Appendix C of \cite{cv}. However, for reasons explained in the
next sub-section, construction of the quantum Hamiltonian operator
based on (\ref{mubar}) was thought not to be feasible. Therefore,
this avenue was used only to gain qualitative insights and was not
pursued in the full quantum theory.

\subsection{The quantum Hamiltonian constraint}
\label{s3.3}

With the curvature operator $\hat{F}_{ab}{}^k$ at hand, it is
straightforward to construct the quantum analog of the Hamiltonian
constraint (\ref{ham1}) because the triad operators can be readily
constructed from the three $\hat{p}_i$. Ignoring for a moment the
factor-ordering issues, the gravitational part of this operator is
given by
\begin{align} \label{qham1} \hat{\mathcal{C}}_{\g}=& -\f{1}
{8\pi G\gamma^2\Delta \lp^2} \,
\left[p_1p_2|p_3|\sin\bar\mu_1c_1\sin\bar\mu_2c_2
+p_1|p_2|p_3\sin\bar\mu_1c_1\sin\bar\mu_3c_3 \right.
\nonumber\\&\qquad\qquad\qquad \left.+|p_1|p_2p_3
\sin\bar\mu_2c_2\sin\bar\mu_3c_3 \right] \end{align}
where for simplicity of notation here and in what follows we have
dropped hats on $p_i$ and $\sin \bar\mu_ic_i$. To write the action
of this operator on $\Hkg$, it suffices to specify the action of
the operators $\exp(i\bar\mu_ic_i)$ on the kinematical states
$\Psi(p_1, p_2, p_3)$. The expression (\ref{mubar}) of $\bar\mu_i$
and the Poisson brackets (\ref{pb}) imply:
\be \label{exp} \exp(\pm i\bar\mu_1c_1) =\exp \Big(\mp 8\pi \gamma
\,\sqrt{\Delta}\,\lp^3\,\,\sqrt{\left|\f{p_1}{p_2p_3}\right|}\,\,\f{\dd}{\dd
p_1}\Big)\, \ee
and its cyclic permutations. At first sight this expression seems
too complicated to yield a manageable Hamiltonian constraint.\\

\begin{itemize}
\item {\emph{Remark:}\, In the isotropic case, the corresponding
expression is simply
$$  \exp(\pm i\bar\mu c) =\exp \Big(\mp 8\pi \gamma
\,\sqrt{\Delta}\,\,\sqrt{\left|\f{1}{p}\right|}\f{\dd} {\dd
p}\Big)\, .$$
Since $\textstyle{1\over\sqrt{p}}\, \f{\dd}{\dd p} \sim
\f{\dd}{\dd v}$, where $v \sim |p|^{3/2}$ is the physical volume
of the fiducial cell $\V$, this operator can be essentially
written as $\exp (\dd/\dd v)$ and acts just as a displacement
operator on functions $\Psi(v)$ of $v$. In the operator
(\ref{exp}) by contrast, all three $p_i$ feature in the exponent.
This is why its action was deemed unmanageable. As we noted at the
end of section \ref{s3.2}, progress was made \cite{chiou,cv} by
simply postulating an alternative, more manageable expression
$\bar{\mu}_i = (\sqrt{\Delta}\,\lp/\sqrt{|p_i|})$, the obvious
analog of $\bar\mu = (\sqrt{\Delta}\,\lp)/\sqrt{|p|}$ in the
isotropic case \cite{aps3}. Then each $\exp (\pm i \bar\mu_ic_i)$
can be expressed essentially as a displacement operator $\exp
\dd/\dd v_i$ with $v_i \sim |p_i|^{3/2}$ and the procedure used in
the isotropic case could be implemented on states
$\Psi(v_1,v_2,v_3)$. Bianchi I quantum dynamics then resembled
three copies of the isotropic dynamics. However, as noted above
this solution is not viable \cite{szulc}.}
\end{itemize}

Our new observation is that the operator (\ref{exp}) can in fact
be handled in a manageable fashion. Let us first make an algebraic
simplification by introducing new dimensionless variables
$\l_i$\,:
\be \l_i\, =\, \f{\sgn(p_i)\,\sqrt{|p_i|}}{(4\pi |\gamma|\,
\sqrt\Delta\, \lp^3)^{1/3}}\, , \ee
(so that $\sgn(\l_i)=\sgn(p_i)$). Then, we can introduce a new
orthonormal basis $|\l_1, \l_2,\l_3\rangle$ in $\Hkg$ by an
obvious rescaling. These vectors are again eigenvectors of the
operators $p_i$\,:
\be p_i|\l_1, \l_2,\l_3\rangle\, =\, \sgn(\l_i)\,
(4\pi|\gamma|\sqrt{\Delta}\,\lp^3)^{\f{2}{3}}\, \l_i^2\, |\l_1,
\l_2,\l_3\rangle \, .\ee
We can expand out any ket $|\Psi\rangle$ in $\Hkg$ as
$|\Psi\rangle = \Psi(\l_1,\l_2,\l_3)\,|\l_1, \l_2,\l_3\rangle$ and
re-express the right side of (\ref{exp}) as an operator on wave
functions $\Psi(\vec\l)$,
\be \exp(\pm i\bar\mu_1c_1) =
\exp\left(\f{\mp\,\sgn(\l_1)}{\l_2\l_3}\,\f{\dd}{\dd\l_1} \right) =:
E_1^\mp\, , \ee
where the notation $E_i^\pm$ has been introduced as shorthand.
(Here, we have used the property $\gamma = \sgn (p_1p_2p_3)
|\gamma|$ of the Barbero-Immirzi parameter from Appendix
\ref{a1}.) To obtain the explicit action of $E_i^\pm$ on wave
functions $\Psi(\vec{\l})$ we note that, since the operator is an
exponential of a vector field, its action is simply to drag the
wave function $\Psi(\vec{\l})$ a unit affine parameter along its
integral curves. Furthermore, since the vector field $\dd/\dd\l_1$
is in the $\l_1$ direction, the coefficient $1/\l_2\l_3$ is
constant along each of its integral curves. Therefore it is
possible to write down the explicit expression of $E_i^\pm$:
\be \label{E} \Big(E_1^\pm\, \Psi\Big)\, \big(\l_1,\l_2,\l_3\big)\,
=\, \Psi \big(\l_1\pm\f{\sgn(\l_1)}{\l_2\l_3}, \l_2,\l_3\big)\, .
\ee
The non-triviality of this action lies in the fact that while the
wave function is dragged along the $\l_1$ direction, the
\emph{affine distance involved in this dragging depends on
$\l_2,\l_3$.} This operator is well-defined because our states
have support only on a countable number of $\l_i$. In particular,
the image $\big(E^\pm_1\, \Psi\big)(\vec\l)$ vanishes identically
at points $\l_2=0$ or $\l_3=0$ because $\Psi$ does not have
support at $\l_1 = \infty$. Thus the factor $\lambda_2\lambda_3$
appearing in the denominator does not cause difficulties.

We can now write out the gravitational part of the Hamiltonian
constraint:
\be \label{qham2} \hat{\mathcal{C}}_\g =
\hat{\mathcal{C}}_{\g}^{(1)}+
\hat{\mathcal{C}}_{\g}^{(2)}+\hat{\mathcal{C}}_{\g}^{(3)}, \ee
with
\ba \label{qham3} \hat{\mathcal{C}}_{\g}^{(1)}=& - \pi \hbar\lp^2
\sqrt{|\l_1\l_2\l_3|}\,\,\Big[\sin\bar\mu_2c_2\, \sgn \l_2\,
|\l_1\l_2\l_3|\, \sgn \l_3\sin\bar\mu_3c_3 \nonumber \\
&+ \sin\bar\mu_3c_3\, \sgn \l_3\, |\l_1\l_2\l_3|\, \sgn
\l_2\sin\bar\mu_2c_2\,\Big]\,\, \sqrt{|\l_1\l_2\l_3|}\ea
where we have used the simplest symmetric factor ordering that
reduces to the one used in \cite{acs} in the isotropic case. (
$\hat{\mathcal{C}}_{\g}^{(2)}$ and $\hat{\mathcal{C}}_{\g}^{(3)}$
are given by the obvious cyclic permutations.) In Appendix \ref{a1},
we show that, under the action of reflections $\hat\Pi_i$ on $\Hkg$,
the operators $\sin\bar\mu_ic_i$ have the same transformation
properties that $c_i$ have under reflections $\Pi_i$ in the
classical theory. As a consequence, $\hat{\mathcal{C}}_\g$ is also
reflection symmetric. Therefore, its action is well defined on
$\Hkg$: $\hat{\mathcal{C}}_\g$ is a densely defined, symmetric
operator on this Hilbert space. In the isotropic case, its analog
has been shown to be essentially self-adjoint \cite{warsaw2}. In
what follows we will assume that (\ref{qham2}) is essentially
self-adjoint on $\Hkg$ and work with its self-adjoint extension.

Finally, it is straightforward to write down the quantum analog of
the full Hamiltonian constraint (\ref{ham1}):
\be \label{qham4} -\hbar^2 \partial^2_\T \, \Psi(\vec{\l}, \T) =
\Theta\, \Psi(\vec{\l}, \T)\ee
where $\Theta = -\mathcal{C}_\g$. As in the isotropic case, one
can obtain the physical Hilbert space $\Hp$ by a group averaging
procedure and the result is completely analogous. Elements of
$\Hp$ consist of `positive frequency' solutions to (\ref{qham4}),
i.e., solutions to
\be \label{qham5} -i\hbar \partial_T \Psi(\vec{\l}, \T)\, = \,
\sqrt{|\Theta|}\, \Psi(\vec{\l}, \T)\, ,\ee
which are symmetric under the three reflection maps $\hat\Pi_i$,
i.e. satisfy
\be \label{sym} \Psi(\l_1,\l_2,\l_3,\, \T) =
\Psi(|\l_1|,|\l_2|,|\l_3|,\, \T)\, . \ee
The scalar product is given simply by:
\ba \label{ip1} \langle \Psi_1|\Psi_2\rangle &=& \langle
\Psi_1(\vec{\l}, \T_o)|\Psi_2(\vec{\l}, \T_o) \rangle_{\rm kin}
\nonumber\\
&=& \sum_{\l_1,\l_2,\l_3} \bar\Psi_1(\vec{\l}, \T_o)\,
\Psi_2(\vec{\l}, \T_o) \ea
where $\T_o$ is any ``instant'' of internal time $\T$.

\begin{itemize}
\item \emph{Remark}: In the isotropic LQC literature
    \cite{aps3,apsv,warsaw1} one began in the classical theory
    with proper time $t$ (which corresponds to the lapse
    function $N_{(t)}=1$) and made a transition to the
    relational time provided by the scalar field only in the
    construction of the physical sector of the quantum theory.
    If we had used that procedure here, the factor ordering of
    the Hamiltonian constraint would have been slightly
    different. In this paper, we started out with the lapse $N =
    |p_1p_2p_3|^{1/2}$ already in the classical theory because
    the resulting quantum Hamiltonian constraint is simpler. In
    the isotropic case, for example, this procedure leads to an
    \emph{analytically soluble} model (the one obtained in
    \cite{acs} by first starting out with $N_{(t)} =1$, then
    going to quantum theory, and finally making some
    well-motivated but simplifying assumptions). It also has
    some conceptual advantages because it avoids the use of
    ``inverse scale factors'' altogether.

\end{itemize}
\subsection{Simplification of $\hat{\mathcal{C}}_\g$}
\label{s3.4}

It is straightforward to expand out the Hamiltonian constraint
$\hat{\mathcal{C}}_\g$ using the explicit action of operators
$\sin (\bar\mu_ic_i)$ given by (\ref{E}) and express it as a
linear combination of 24 terms of the type
\be \label{cij} \hat{\mathcal{C}}_{ij}^{\pm\,\pm} := \sqrt{|v|}\,
E^\pm_i\sgn(\l_i) \,|v|\, \sgn(\l_j) E^\pm_j\, \sqrt{|v|}\, , \ee
(where, $i\not=j$ and as before there is no summation over $i,j$).
Unfortunately, the $\sgn (\l_i)$ factors in this expression and in
the action of $E^\pm_i$ make the result quite complicated. More
importantly, it is rather difficult to interpret the resulting
operator. The expression can be simplified if we introduce the
volume of $\V$ as one of the arguments of the wave function. In
particular, this would make quantum dynamics easier to compare
with that of the Friedmann models. With this motivation, let us
further re-arrange the configuration variables and set
\be \label{v} v = 2\,\l_1\l_2\l_3\, . \ee
The factor of $2$ in (\ref{v}) ensures that this $v$ reduces to
the $v$ used in the isotropic analysis of \cite{aps3} (if one uses
the value of $\Delta$ used there). As the notation suggests, $v$
is directly related to the volume of the elementary cell $\V$:
\be \hat{V}\, \Psi (\l_1,\l_2,v) = 2\pi\, |\gamma|\, \sqrt\Delta\,
|v|\, \lp^3\, \Psi (\l_1,\l_2,v)\, . \ee
 One's first impulse would be to introduce two
other variables in a symmetric fashion, e.g., following Misner
\cite{misner}. Unfortunately, detailed examination shows that they
make the constraint (\ref{qham2}) even less transparent!%
\footnote{Misner-like variables ---volume and logarithms of metric
components--- were used in the brief discussion of Bianchi I
models in \cite{mb-stable}. This discussion already recognized
that the use of volume as one of the arguments of the wave
function would lead to simplifications. Dynamics was obtained by
starting with the Hamiltonian constraint in the $\mu_o$ scheme
from \cite{mbBI} and then substituting $\bar\mu_i$ of
(\ref{mubar}) for $\mu_o^i$ in the final result. This procedure
does simplify the leading order quantum corrections to dynamics.
By contrast, our goal is to simplify the full constraint. More
importantly, constraint (\ref{qham2}) is an improvement over that
of \cite{mb-stable} because we introduced $\bar\mu_i$ from the
beginning of the quantization procedure and systematically defined
the operators $\sin (\bar\mu_ic_i)$ (in section \ref{s3.3}).}

Let us simply use $\lambda_1, \lambda_2, v$ as the configuration
variables in place of $\lambda_1,\lambda_2,\lambda_3$. This change
of variables would be non-trivial in the Schr\"odinger
representation but is completely tame here because the norms on
$\Hkg$ are defined using a discrete measure on $\R^3$. As a
consequence, the scalar product is again given by the sum in
(\ref{ip1}), the only difference is that $\l_3$ is now replaced by
$v$. Since the choice $(\l_1,\l_2,v)$ breaks the permutation
symmetry, one might have first thought that it would not be
appropriate. Somewhat surprisingly, as we will now show, it
suffices to make the structure of the constraint transparent. (Of
course, the simplification of the constraint would have persisted
if we had chosen to replace either $\l_1$ or $\l_2$
---rather than $\l_3$--- with $v$.) Finally, note that the positive
octant is now given by $\l_1\ge 0, \l_2\ge 0$ and $v \ge 0$.

To obtain the explicit action of the constraint, it is extremely
convenient to use the fact that states $\Psi$ in $\Hkg$ satisfy the
symmetry condition (\ref{sym}) and that $\hat{\mathcal{C}}_\g$ has a
well defined action on this space. Therefore, to specify its action
on any given $\Psi$ it suffices to find the restriction of the image
$\Phi(\l_1,\l_2,v) := (\hat{\mathcal{C}}_\g\,\Psi)(\l_1,\l_2,v)$ to
the positive octant. The value of $\Phi$ in other octants is
determined by its symmetry property. \emph{This fact greatly
simplifies our task} because we can use it to eliminate the
$\sgn(\l_i)$ factors in various terms which complicate the
expression tremendously.

For concreteness let us focus on one term in the constraint operator
(which turns out to be the most non-trivial one for our
simplification):
\ba &&\Big(\hat{\mathcal{C}}_{21}^{--}\, \Psi\Big) (\l_1,\l_2,v):=
\Big(\sqrt{|v|}\,E^-_2\sgn(\l_2)\,
|v|\,\sgn(\l_1)\, E^-_1 \sqrt{|v|}\, \Psi\Big)\, (\l_1,\l_2, v)\nonumber\\
&=& \big[\sqrt{|v|}\, \sgn(\l_2(1-\f{2\sgn\l_2}{v}))\,\, |v-2\sgn
\l_2|\,\, \sgn (\l_1)\, \sqrt{|v-2\sgn \l_1 - 2\sgn
\l_2|}\,\big]\times \, \nonumber\\
& & \quad \Psi\Big(\f{v-2\sgn\l_1-2\sgn\l_2}{v-2\sgn\l_2}\l_1,\,\,
\f{v-2\sgn\l_2}{v}\l_2,\,\, v-2\sgn\l_1-\sgn\l_2\Big)\, .\ea
If we now restrict the argument of
$\big(\hat{\mathcal{C}}_{12}^{--}\, \Psi\big)$ to the positive
octant, the expression simplifies:
\be \label{simple} \Big(\hat{\mathcal{C}}_{21}^{--}\,
\Psi\Big)\Big|_{\rm +\,octant} = \big[\sqrt{v}(v-2)
\sqrt{|v-4|}\,\big]\,\, \Psi \big(\f{v-4}{v-2}\l_1,\,
\f{v-2}{v}\l_2,\, v-4\big)\, . \ee
Now the action of this operator is more transparent: the wave
function is multiplied by functions \emph{only} of volume and, in
the argument of the wave function, volume simply shifts by -4 and
$\l_1, \l_2$ are rescaled by multiplicative factors which also
depend \emph{only}  on the volume. Since the full constraint is a
linear combination of terms of this form, its action is also
driven primarily by volume. As we will see, this key property
makes the constraint manageable and greatly simplified the task of
analyzing the relation between the LQC quantum dynamics of Bianchi
I and Friedmann Models. From now on, unless otherwise stated,
\emph{we will restrict the argument of the images
$\big(\hat{\mathcal{C}}_{ij}^{\pm}\, \Psi\Big)$ to lie in the
positive octant}; its value in other octants is given simply by
$\big(\hat{\mathcal{C}}_{ij}^{\pm\,\pm}\, \Psi\big)(\l_1, \l_2, v)
= \big(\hat{\mathcal{C}}_{ij}^{\pm\,\pm}\, \Psi\big)(|\l_1|,
|\l_2|, |v|)$.

The form (\ref{simple}) of the action of operators
$\hat{\mathcal{C}}_{ij}^{\pm\,\pm}\,$ enables us to discuss
singularity resolution. For completeness, let us first write out
the four terms corresponding to $i,j$=$1,2$ (which are the most
complicated of the 24 terms in $\hat{\mathcal C}_\g$):
\begin{align} \label{ops1}
\Big(\hat{\mathcal{C}}_{21}^{++}\Psi\Big)\,(\l_1,\l_2,v) &= (v+2)
\sqrt{v(v+4)}\cdot\Psi\big(\f{v+4}{v+2}\l_1,\f{v+2}
{v}\l_2,v+4\big), \\
\Big(\hat{\mathcal{C}}_{21}^{+-}\Psi\Big)\, (\l_1,\l_2,v)
&= v(v+2)\cdot\Psi\big(\f{v}{v+2}\l_1,\f{v+2}{v}\l_2,v\big), \\
\Big(\hat{\mathcal{C}}_{21}^{-+} \Psi\Big)(\l_1,\l_2,v)
&= v(v-2)\cdot\Psi\big(\f{v}{v-2}\l_1,\f{v-2}{v}\l_2,v\big), \\
\label{ops4}\Big(\hat{\mathcal{C}}_{21}^{--} \Psi\Big)\,(\l_1,\l_2,v)
&=(v-2)\sqrt{v|v-4|}\cdot
\Psi\big(\f{v-4}{v-2}\l_1,\f{v-2}{v}\l_2,v-4\big)\, . \end{align}
Recall that, since $v$ is proportional to the volume of the
elementary cell, it vanishes when any one of the three directional
scale factors $a_i$ vanish. Thus, the classical singularity
corresponds precisely to the points at which $v$ vanishes. Now
suppose that the function $\Psi(\l_1,\l_2,v)$ has no support on
points $v=0$ at an initial internal time $\T_o$. As it evolves via
(\ref{qham4}), can it end up having support on such points? We will
argue that this is impossible.

Let us decompose $\Hkg$ as $\Hkg = \H_{\rm sing}^{\g} \oplus \H_{\rm
reg}^{\g}$ where $\Psi(\l_1,\l_2,v)$ is in $\H_{\rm sing}^{\g}$ if
it has support only on points with $v=0$ and it is in $\H_{\rm
reg}^{\g}$ if it has no support on points with $v=0$. Now, all the
operators $\hat{\mathcal{C}}_{ij}^{\pm\,\pm}$ have a factor of
$\sqrt{v}$ acting on the right (see Eq. (\ref{cij})). It ensures
that each $\hat{\mathcal{C}}_{ij}^{\pm\,\pm}$ annihilates every
state in $\H_{\rm sing}^{\g}$. Therefore $\H_{\rm sing}^{\g}$ is
left invariant by the evolution. More importantly, because of the
pre-factors of $v\pm 2$ and $v\pm 4$ the action of the 4 operators
in (\ref{ops1}) - (\ref{ops4}) preserves $\H_{\rm reg}^{\g}$. This
property is shared also by $\hat{\mathcal{C}}_{ij}^{\pm\,\pm}$ for
other values of $i,j$ and hence by $\hat{\mathcal{C}}_\g$ and all
its powers.%
\footnote{To make this argument mathematically rigorous one would
have to establish that $\hat{\mathcal{C}}_\g$ is essentially
self-adjoint and its self adjoint extension also shares this
property (or a suitable generalization thereof).}
Therefore, the relational dynamics of (\ref{qham4}) decouples
$\H_{\rm sing}^{\g}$ from $\H_{\rm reg}^{\g}$. In particular, if one
starts out with a ``regular'' quantum state at $\T=0$, it remains
regular throughout the evolution. In this precise sense, the
singularity is resolved.

Next, let us write out explicitly the full Hamiltonian constraint
(\ref{qham4}):
\begin{align} \label{qham6} \partial_\T^2\, \Psi(\l_1,\l_2,v;\T) =& \f{\pi G}
{2}\sqrt{v}\Big[(v+2)\sqrt{v+4}\,\Psi^+_4(\l_1,\l_2,v;\T) - (v+2)\sqrt
v\, \Psi^+_0( \l_1,\l_2,v;\T)\nonumber \\& -(v-2)\sqrt v\,
\Psi^-_0(\l_1,\l_2,v;\T) + (v-2)
\sqrt{|v-4|}\,\Psi^-_4(\l_1,\l_2,v;\T)\Big], \end{align}
where $\Psi^\pm_{0,4}$ are defined as follows:
\begin{align} \label{qham7}\Psi^\pm_4(\l_1,\l_2,v;\T)=& \:\Psi
\left(\f{v\pm4}{v\pm2}\cdot\l_1,\f{v\pm2}{v}\cdot\l_2,
v\pm4;\T\right)+\Psi\left(\f{v\pm4}{v\pm2}\cdot\l_1,\l_2,
v\pm4;\T\right)\nonumber\\& +\Psi\left(\f{v\pm2}{v}\cdot\l_1,
\f{v\pm4}{v\pm2}\cdot\l_2,v\pm4;\T\right)+\Psi
\left(\f{v\pm2}{v}\cdot\l_1, \l_2,v\pm4;\T\right)\nonumber
\\&+\Psi\left(\l_1,\f{v\pm2}{v}\cdot\l_2,
v\pm4;\T\right)+\Psi\left(\l_1,\f{v\pm4}{v\pm2}\cdot\l_2,v\pm4;\T\right),
\end{align}
and
\begin{align} \label{qham8} \Psi^\pm_0(\l_1,\l_2,v;\T)=
& \:\Psi\left(\f{v\pm2}{v}\cdot\l_1, \f{v}{v\pm2}\cdot\l_2,v;\T\right)
+\Psi\left(\f{v\pm2}{v}\cdot\l_1,\l_2,v;\T \right)\nonumber \\&
+\Psi\left(\f{v}{v\pm2}\cdot\l_1,\f{v\pm2}{v}\cdot\l_2,v;
\T\right)+\Psi\left(\f{v}{v\pm2}\cdot\l_1,\l_2,v;\T\right)\nonumber
\\& +
\Psi\left(\l_1,\f{v}{v\pm2}\cdot\l_2,v;\T\right)+\Psi\left(\l_1,\f{v\pm2}{v}
\cdot\l_2,v;\T\right)\, ,\end{align}
where, as before, we have given the restriction of the image of
$\hat{\mathcal{C}}_\g$ to the positive octant. Because $\H^\g_{\rm
reg}$ is left invariant by evolution we can in fact restrict
$\lambda_1,\lambda_2,v$ to be strictly positive. On the right sides
of (\ref{qham7}) and (\ref{qham8}), arguments of $\Psi$ can take
negative values. However, since $\Psi(\l_1,\l_2,v) = \Psi(|\l_1|,
|\l_2|, |v|)$, \emph{we can just introduce absolute value signs on
these arguments}. Consequently, knowing the restriction of $\Psi$ to
the positive octant, (\ref{qham7}) and (\ref{qham8}) enable us to
directly calculate its image under $\hat{\mathcal{C}}_\g$. In
particular, numerical evolutions can be carried out by restricting
oneself to the positive octant.

Let us now examine the structure of this equation. As in the
isotropic case, the right side is a difference equation. As far as
the $v$ dependence is concerned, the steps are uniform: the argument
of the wave function involves $v-4,\,v,\,v+4$ exactly as in the
isotropic case. The step sizes are also the same as in \cite{aps3}
because, as noted above, our variable $v$ is in precise agreement
with that used in the isotropic case. There is again superselection.
For each $\epsilon \in [0,4)$, let us introduce a `lattice'
$\mathcal{L}_\epsilon$ consisting of points
$v= 4n$ if $\epsilon=0$ and $v= 2n+\epsilon$ if $\epsilon\not=0$.%
\footnote{As in the isotropic case, the lattice is doubled if
$\epsilon\not=0\,\,{\rm or}\,\,2$ because of the symmetry property
of our wave functions.}
Then the quantum evolution ---as well as the action of the Dirac
observables--- preserves the subspaces $\Hp^{\epsilon}$ consisting
of states with $v$-support on $\mathcal{L}_\epsilon$. The most
interesting of these sectors is the one labelled by $\epsilon=0$
since it contains the classically singular points, $v=0$.
\emph{Therefore in what follows, unless otherwise stated, we will
restrict ourselves to this sector.}

The dependence of $\hat{\mathcal{C}}_\g\, \Psi$ on $\lambda_1,
\lambda_2$, by contrast, is much more difficult to control
technically because the first two arguments of the wave function
cannot be chosen to lie on a regular lattice in any simple way. In
particular, even if we started out with a wave function which has
support only on a lattice, say $\l_1 = n\l_o$ for some $\l_o$, the
action of $\hat{\mathcal{C}}_\g$ shifts support to points such as
$\lambda_1 = [(v\pm 2)/v] n\l_o$ which do not lie on this lattice.
Thus, there is no obvious superselection with respect to $\l_1$
and $\l_2$; we have to work with the entire $\R^2$ they span. Had
it been permissible to set  $\bar\mu_i \propto /\sqrt{|p_i|}$, we
could have restricted $\l_i$ to lie on a regular lattice
\cite{chiou}. Then, following \cite{szulc}, we could have repeated
the strategy used successfully in the isotropic case in \cite{acs}
to simplify dynamics by carrying out a Fourier transform to pass
to variables which are conjugate to $\l_1, \l_2$. However, as
remarked earlier, that choice of $\bar\mu_i$ is inadmissible and
hence the strategy cannot be repeated in the Bianchi I case.
Nonetheless, it is still feasible to carry out numerical
simulations. For, if one knows the support of the quantum state at
an initial time $\T_o$ and the number of time-steps across which
one wants to evolve, one can calculate the number of points on a
(irregular) grid in the $\l_1$-$\l_2$ plane on which the wave
function will have support. Numerical work has in fact already
commenced \cite{ht}. It would be interesting to investigate
whether the efficient algorithms that have been introduced in the
context of regular lattices \cite{sk} can be extended to this
case.

We will conclude this discussion by noting that it is possible to
read off some qualitative features of dynamics from (\ref{qham6}) --
(\ref{qham8}). Since the steps in $v$ of this difference equation
are the same as those in the isotropic case, the dynamics of volume
---and also of the matter density $\hat\rho_\m$, since
$\hat{p}_{(\T)}$ is a constant of motion--- would be qualitatively
similar to that in the isotropic case. What about anisotropies?
The $\l_I$ ($I=1,2$) do not feature in the overall numerical
factors in (\ref{qham6}); they appear only in the argument of the
wave functions. Under the action of $\hat{\mathcal{C}}_\g$, these
arguments get rescaled by factors $v\pm 4/v\pm 2, \, v\pm2/v$ and
$v/v\pm2$. For large volumes, or more precisely low densities,
these factors go as $1 + O(\rho_\m/\rp)$. Hence, to leading order,
we will recover of the classical result that $a_1a_2a_3 (H_i -
H_j)$ are constants, where $a_i$ are the directional scale factors
and $H_i := \dd \ln a_i/\dd t$, the directional Hubble parameters.
Since quantum corrections go as $\rho/\rp$ they are utterly
negligible away from the Planck regime.

In the next section we will discuss three important features of
dynamics dictated by (\ref{qham6}) which provide significant
physical intuition in complementary directions.

\section{Properties of the LQC quantum dynamics}
\label{s4}

This section is divided into three parts. Since we have used the
same general procedure as in the isotropic case it is natural to ask
how the quantum dynamics of (\ref{qham6}) compares to that in
\cite{aps3}. In the first part we show that there is a natural
projection from a dense subspace of the physical Hilbert space of
the Bianchi I model to that of the Friedmann model which maps the
Bianchi I Hamiltonian constraint to that of the Friedmann model.
This result boosts confidence in the overall coherence and
reliability of the quantization scheme used in LQC. In various
isotropic models \cite{aps3,apsv,kv2,bp,ap}, one can derive certain
effective equations. Somewhat surprisingly, for states which are
semi-classical at a late initial time, they faithfully capture
quantum dynamics throughout the entire evolution, including the
bounce. The same considerations lead to effective equations in
Bianchi I models which were already analyzed by Chiou and
Vandersloot in Appendix C of \cite{cv}. In the second sub-section we
briefly discuss these equations and their consequences. In the
third, we show that, as in the isotropic case \cite{aps3,acs}, there
is a precise sense in which the LQC quantum dynamics reduces to that
of the Wheeler-DeWitt theory in the low curvature regime.

\subsection{Relation to the LQC Friedmann dynamics}
\label{s4.1}

The problem of comparing dynamics of a more general system with that
of a restricted, symmetry reduced one has been discussed in the
literature in several contexts. In the classical theory, symmetric
states often provide symplectic sub-manifolds $\Gamma_\res$ of the
more general phase spaces $\Gamma_\gen$. Furthermore $\Gamma_\res$
are preserved by the dynamics on $\Gamma_\gen$. Therefore, it is
tempting to repeat the same strategy in the quantum theory. Indeed,
sometimes it is possible to find natural sub-spaces $\H_\res$ of
states with additional symmetry in the full Hilbert space $\H_\gen$
of the more general system. However, generically $\H_\res$ is not
left invariant by the more general dynamics (see, e.g.,
\cite{kr,rt}). In our case, one can introduce an isotropic sub-space
of $\H_\res$ in the quantum theory based on any given fiducial cell
$\V$: isotropic states correspond to wave functions
$\Psi(\l_1,\l_2,v)$ which have support only at points $\l_1 = \l_2 =
(v/2)^{1/3}$. (But note that this sub-space is not invariantly
defined; it is tied to $\V$!) It is easy to check that the space
$\H_\res$ of these states is not left invariant by the Bianchi I
quantum dynamics (\ref{qham6}).

However, this fact cannot be interpreted as saying that there is no
simple relation between the quantum dynamics of the two theories:
since restriction to $\H_\res$ amounts to a sharp freezing of
anisotropic degrees of freedom, in view of the quantum uncertainty
principle, this procedure is not well suited to compare the quantum
dynamics of the two systems. As pointed out in section \ref{s1}, a
better strategy is to integrate out the extra, anisotropic degrees
of freedom. This would correspond to a \emph{projection map} from
$\H_\gen$ to $\H_\res$ rather than an embedding of $\H_\res$ into
$\H_\gen$.

Consider first, as an elementary example, a particle moving in
$\R^3$. Suppose that the potential depends only on $z$ so that
dynamics has a symmetry in the $x,y$ directions. In the classical
theory, there are several natural embeddings of the phase space
$\Gamma_\res$ into $\Gamma_\gen$. For example, we can set $(z,p_z)
\rightarrow$ ($x$=$x_o$, $y$=$y_o$,$z$;\, $p_x$=0, $p_y$=0, $p_z$)
and the Hamiltonian vector field of the full theory is then
tangential to the images of each of these embeddings. However, in
the quantum theory the Hilbert space $\H_\gen$ of the full system is
$L^2(\R^3, \dd^3x)$ and there is no natural embedding $\psi(z)
\rightarrow \Psi(x,y,z)$. The classical strategy would suggest
setting $\Psi(x,y,z) = \delta(x, x_o)\, \delta(y,y_o) \psi(z)$ but
this is not a normalizable state in $\H_\gen$ for any $\psi(z)$.
Even if one were to ignore this fact and try to evolve these states,
one would find that they are not preserved by the full Hamiltonian
operator $\hat{H}$.

Note however that there \emph{is} a natural projection $\P$ from a
dense subspace in $\H_\gen$ to that in $\H_\res$:
\be \Psi(x,y,z)\, \rightarrow\, (\P\Psi)(z) := \int dx\int dy\,
\Psi(x,y,z)\, \equiv\, \psi(z) \, . \ee
(For example, we can choose the dense subspace to be the space of
smooth functions of compact support.) Furthermore, under this
projection, the Hamiltonian operator
$$\hat{H} = -(\hbar^2/2m) \Delta + V(z)$$
of the general system is mapped to the Hamiltonian operator
$$\hat{h} := -(\hbar^2/2m) \dd^2/\dd z^2 + V(z)$$
of the reduced system. Hence solutions $\Psi(\vec{x}, t)$ of the
Schr\"odinger equation of the full system are mapped to solutions
$\psi(z,t)$ of the reduced system. Finally, this projection
strategy continues to work for more general Hamiltonians of the
type $f^i(z)p_i + V(z)$ which again have a symmetry in the $x,y$
directions.

Let us return to the Bianchi I model and define a projection $\P$
from states $\Psi(\l_1,\l_2,v)$ of the Bianchi I model to the states
$\psi(v)$ of the Friedmann model of \cite{aps3} as follows:
\be \label{proj} \Psi(\l_1, \l_2, v)\, \rightarrow\, (\P\Psi)(v) :=
\sum_{\l_1,\l_2}\, \Psi (\l_1,\l_2,v) \equiv \psi(v)\, .\ee
(The idea of using such a map already appeared in \cite{bhm} where
the map was defined between elements of ${\rm Cyl}^\star$ of the
locally rotationally symmetric Bianchi I model and that of the
Friedmann model.) Again, $\P$ is a well defined projection from a
dense subspace of the Bianchi I Hilbert space to a dense subspace of
the Friedmann Hilbert space, consisting, for example, of states
which have support only on a finite number of points. As is manifest
from (\ref{proj}), its effect is to focus on volume by ``integrating
out'' the anisotropic degrees of freedom with the same volume.
Applying this projection map $\P$ to Eq. (\ref{qham6}), we find
\begin{align} \label{projham} \partial_\T^2 \psi(v;\T)\, =\,& 3\pi G\Big[(v+2)
\sqrt{v(v+4)}\psi(v+4;\T)-2v^2\psi(v;\T) \nonumber \\&
\qquad\qquad +(v-2) \sqrt{v(v-4)}\psi(v-4;\T)\Big]. \end{align}
This is \emph{precisely} the quantum constraint describing the LQC
dynamics of the Friedmann model with lapse%
\footnote{As noted at the end of section \ref{s3.3}, the analysis
in \cite{aps3} began with the lapse $N=1$ and therefore leads to a
slightly different factor ordering. Had one used $N=|p|^{3/2}$
from the beginning as in the current paper, one would have
obtained the factor ordering used in \cite{acs}. Eq.
(\ref{projham}) matches exactly with that constraint.}
$N= {|p|}^{3/2}$. The reason for the exact agreement is two-fold.
First, the Hamiltonian constraint $\hat{\mathcal{C}}_\g$ of the
Bianchi I model is a difference operator whose coefficients depend
\emph{only} on $v$ and, second, the shift in the argument is
dictated \emph{only} by $v$. Thus, conceptually, $\l_1,\l_2$ are
``inert directions'' in the same sense that $x,y$ are in the
elementary example discussed above. To summarize, there is a
simple ---and \emph{exact}--- relation between quantum dynamics of
the two theories. It would be interesting to investigate if this
result admits a suitable extension to other Bianchi models
\cite{awe3,ahs}.

In completely general situations, of course, this exact agreement
will not persist: the projected dynamics will provide extremely
non-trivial corrections to the dynamics of the simpler system.
However, the BKL conjecture says that the dynamics of general
relativity greatly simplifies near space-like singularities: In
this regime, the time evolution at any one spatial point is well
modelled by that of Bianchi I cosmology. Therefore, in a large
class of situations there may well be a sense in which the quantum
dynamics in the deep Planck regime can be projected to that of the
Friedmann model with only small corrections. If so, the Planck
scale quantum dynamics of the isotropic, homogeneous degree of
freedom in the full theory will be much simpler than what one
would have a priori expected.

\subsection{Effective equations}
\label{s4.2}

Physically, the most interesting quantum states are those that are
sharply peaked at a classical trajectory at late times. As
explained in section \ref{s1}, in the isotropic case such states
remain peaked at certain effective trajectories at \emph{all
times}, including the epoch during which the universe undergoes a
quantum bounce. Thus, even in the deep Planck regime quantum
physics is well captured by a smooth metric although its dynamics
can no longer be approximated by the classical Einstein's
equations and its components now contain large, $\hbar$-dependent
terms. The effective equations obeyed by these geometries were
first derived using ideas from geometrical quantum mechanics
\cite{jw,vt}. However, the assumptions made in these derivations
break down in the deep Planck regime. Therefore a priori there was
no reason to expect these equations to describe quantum dynamics
so well also in the Planck regime. That they do was first shown by
numerical simulations of the exact quantum equations
\cite{aps2,aps3} in the k=0, $\Lambda$=0 case. It was then
realized that this model is in fact exactly soluble
\cite{acs,mb-exact} and the power of the effective equations could
be attributed to this property. However, k=0 models with
\emph{non-zero} cosmological constant and the closed k=1 models do
not appear to be exactly soluble. Yet, numerical solutions of the
exact quantum equations show that the effective equations continue
to capture full quantum dynamics extremely well \cite{apsv,bp,ap}.

New light was shed on this phenomenon by recent work on a path
integral formulation of quantum cosmology \cite{ach}. The idea
here is to return to the original derivation of path integrals due
to Feynman and Hibbs \cite{fh} starting from quantum mechanics. In
the isotropic case, then, the strategy is to \emph{begin} with the
kinematics and dynamics of LQC and then rewrite the transition
amplitudes as path integrals. The resulting framework has several
novel features. First, because the LQC kinematics relies on
quantum geometry, paths that feature in the final integral are
different from what one would have naively expected from the
Wheeler-DeWitt theory. Second, the action that features in the
measure is not the Einstein-Hilbert action but contains
non-trivial quantum corrections. When expressed in the phase space
language, $L = p\dot{q} -H(p,q)$, the ``Hamiltonian'' $H$ turns
out to be precisely the effective Hamiltonian constraint derived
in \cite{jw,vt}, even though this casting of the LQC transition
amplitudes in the path integral language is exact and does not
pre-suppose that we are away from the Planck regime. Now, in the
path integral approach, we have the following general paradigm.
Consider the equations obtained by varying the action that appears
in the path integral. (Generally these are just the classical
equations but in LQC they turn out to be the effective equations
of \cite{aps3,apsv,vt}.) Fix a path representing a solution to
these equations. If the action evaluated along this path is large
compared to $\hbar$ then that solution is a good approximation to
full quantum dynamics. If one applies this idea to isotropic LQC,
one is led to conclude that solutions to the effective equations
of \cite{jw,vt} should be good approximations to full quantum
dynamics also in the k=0, $\Lambda\not=$0 and k=1 cases. This is
precisely what one finds in numerical simulations. Thus, the path
integral approach may well provide a deeper explanation of the
power of effective equations. While such a path integral analysis
is yet to be carried out in detail in the anisotropic case,
because of the situation in the simpler cases it is of
considerable interest to find effective equations and study their
implications.

This task was carried out already by Chiou and Vandersloot in the
Appendix C of \cite{cv}. We will summarize the relevant results and
briefly comment on the general picture that emerges.

Without loss of generality, we can restrict ourselves to the
positive octant. Then the effective Hamiltonian constraint is given
simply by the direct classical analog of (\ref{qham1}):
\be \label{eff-ham1} \pT^2 + {\mathcal{C}}^{\rm eff}_{\g} =0 \ee
where
\be \label{eff-ham2} {\mathcal{C}}_{\g}^{\rm eff}= -\f{p_1p_2p_3}
{8\pi G\gamma^2\Delta} \,\, \big[\sin\bar\mu_1c_1\sin\bar\mu_2c_2
+\sin\bar\mu_2c_2\sin\bar\mu_3c_3
+\sin\bar\mu_3c_3\sin\bar\mu_1c_1 \big].  \ee
Since $\sin x$ is bounded by 1 for all $x$, these equations
immediately imply that the matter density, $\rho_{\m} = \pT^2/2V^2
\equiv \pT^2/2p_1p_2p_3$ can never become greater than the
critical density $\rcr \approx 0.41 \rp$, first found in the
isotropic case \cite{aps3,apsv,kv2,acs,cs}. Since $\rho$ becomes
infinite at the big bang singularity in the classical evolution,
there is a precise sense in which the singularity is resolved in
the effective theory.

Effective equations are obtained via Poisson brackets as in section
\ref{s2} but using (\ref{eff-ham1}) in place of the classical
Hamiltonian constraint. This gives, for example,
\be \label{pdot} \f{dp_1}{d\tau} =
\f{p_1\sqrt{p_1p_2p_3}}{\sqrt\Delta\gamma \lp}\,
\cos(\bar\mu_1c_1)\,\Big(\sin\bar\mu_2c_2+\sin\bar\mu_3c_3\Big),
\ee
and
\begin{align} \f{dc_1}{d\tau} = -\f{p_2p_3}{\Delta\gamma\lp^2}
\Big[&\sin\bar\mu_1c_1
\sin\bar\mu_2c_2+\sin\bar\mu_1c_1\sin\bar\mu_3c_3+\sin\bar\mu_2c_2
\sin\bar\mu_3c_3 \nonumber \\&
+\f{\bar\mu_1c_1}{2}\cos\bar\mu_1c_1\big(
\sin\bar\mu_2c_2+\sin\bar\mu_3c_3\big)-\f{\bar\mu_2c_2}{2}\cos\bar\mu_2c_2
\big(\sin\bar\mu_1c_1+\sin\bar\mu_3c_3\big) \nonumber \\&
-\f{\bar\mu_3c_3}{2}
\cos\bar\mu_3c_3\big(\sin\bar\mu_1c_1+\sin\bar\mu_2c_2\big)\Big].
\end{align}
Equations for $p_2,c_2$ and $p_3,c_3$ are obtained by cyclic
permutations. These effective equations include ``leading order
quantum corrections'' to the classical evolution equations
(\ref{dp1}) and (\ref{dc1}). In any solution, these corrections
become negligible in the distant past and in the distant future. As
we noted in section \ref{s2}, the shear $\Sigma$ defined in Eq.
(\ref{Sigma}) is a constant of motion in the classical theory. This
is no longer the case in the effective theory. However, one can show
that it remains finite throughout the evolution, becomes
approximately constant in the low curvature region both in the
distant past and in the distant future. Furthermore, its value in
the distant future is the same as that in the distant past along any
effective trajectory in the phase space.

Vandersloot (personal communication) has also carried out
numerical integration of these equations. In the isotropic case
each effective trajectory undergoes a quantum bounce when the
matter density $\rho_\m$ achieves a critical value $\rcr \approx
0.41\rp$. As one might expect, now the situation is more
complicated because of the additional degrees of freedom. First,
there are now several distinct ``bounces''. More precisely, in
addition to $\rho_\m$ (or the scalar curvature), we now have to
keep track of the three Hubble rates $H_i$ which directly control
the Weyl curvature. In the backward evolution towards the
classical big bang, Einstein's equations approximate the effective
equations extremely well until the density of one of the $H_i$
enters the Planck regime. Then the quantum corrections start
rising quickly. Their net effect is to dilute the quantity in
question. Once the quantity exits the Planck regime as a result of
this dilution, quantum geometry effects again become negligible.
Thus, as in the isotropic case, one avoids \emph{the
ultraviolet-infrared tension \cite{aa-badhonef} because the
quantum geometry effects are extremely strong in the Planck regime
but die off extremely quickly as the system exits this regime.}
Secondly, the ``volume'' or the ``density bounce'' occurs when the
matter density is lower than $\rcr$. This is not surprising
because what matters is the total energy density and now there is
also a contribution from gravitational waves. Finally, although
there are distinct ``bounces'' for density (or scalar curvature)
and the $H_i$ (or the Weyl curvature invariants), they all occur
near each other in the relational time $\T$.

There are indications that the general scenario provided by
effective equations correctly captures the qualitative features of
the full quantum evolution. However, the arguments are not
conclusive. For conclusive evidence for (or against) this picture,
one needs numerical simulations \cite{ht} of the exact quantum
equations of section \ref{s3.4}, or a detailed, path integral
treatment of the Bianchi I models along the lines of \cite{ach}.

\subsection{Relation to the Wheeler-DeWitt Dynamics}
\label{s4.3}

Quantum dynamics of LQC is governed by a \emph{difference}
---rather than a \emph{differential}--- equation because of the quantum
geometry effects. However, we will now show that, as in the
isotropic case \cite{aps3,apsv,acs}, the LQC quantum dynamics is
well approximated by the Wheeler-DeWitt (WDW) differential
equation away from the Planck regime where quantum geometry
effects become negligible.

In the \WDW theory the directional scale factors, and hence the
three $\l_i$ can assume any real value and it is simpler to work
with the three $\l_i$ rather than with $\l_1,\l_2,\,v=
2\l_1\l_2\l_3$. Let us therefore set $\ul{\Psi}(\l_1,\l_2,\l_3;\,
\T) = \Psi (\l_1,\l_2,v;\, \T)$ and assume that $\ul\Psi$ admits a
smooth extension to all real values of $\l_i$. The idea is to pair
various terms in Eqs. (\ref{qham7}) and (\ref{qham8}) in such a
way so that two of the three arguments of $\ul\Psi$ are the same.
For example, one such pair is
\be \label{pair} \ul{\Psi}\left(\f{v+4}{v+2}\cdot
\l_1,\f{v+2}{v}\cdot\l_2, \l_3;\T \right) \quad \mathrm{and} \quad
\ul\Psi\left(\f{v}{v+2}\cdot\l_1,\f{v+2}{v}\cdot
\l_2,\l_3;\T\right). \ee
Next, let us define $v'=v+2$ and $\l'_2=v'\l_2/(v'-2)$ so that we
have
\be \sqrt{v+4}= \sqrt{v'+2}=
\sqrt{\left(\l_1+\f{1}{\l'_2\l_3}\right)\l'_2\l_3}\, . \ee
Ignoring the common pre-factors in Eqs. (\ref{qham7}) --
(\ref{qham8}), the two paired terms in Eq. (\ref{pair}) can be
expressed as:
\ba
&\sqrt{v'+2}\,\,\ul\Psi\big(\l_1+\f{1}{\l'_2\l_3},\l'_2,\l_3;\T\big)
-\sqrt{v'-2}\,\,\ul\Psi\big(\l_1-\f{1}{\l'_2\l_3},\l'_2,\l_3;\T\big)
\nonumber\\
&= \f{2}{\l'_2\l_3}\,\f{\partial}{\partial\l_1}\,\sqrt{v'}\,\,
\ul\Psi(\l_1,\l'_2,\l_3;\T) + O\big( (\f{1}{\l_2'\l_3})^n\,
\f{\partial^n}{\partial \l_1^n}\,
\sqrt{v'}\,\ul\Psi \big)\nonumber\\
&=\f{4\l_1}{v'}\,\f{\partial}{\partial\l_1}\sqrt{v'}\,\ul\Psi(
\l_1,\l'_2,\l_3;\T) + O\big( (\f{1}{\l_2'\l_3})^n\,
\f{\partial}{\partial\l_1^n}\, \sqrt{v'}\,\ul\Psi \big) \ea
where $n>1$. (Notice that the $v'$ in the denominator in front of
the partial derivative will cancel the $v+2$ pre-factor in Eq.
(\ref{qham6}).) One can suitably pair all terms in (\ref{qham7})
and (\ref{qham8}) and express them as differential operators with
corrections which are small for large values of $\l_i$. Let us
ignore these corrections ---i.e. assume that the $(1/\l_i\l_j)^n
\partial_k^n\, \sqrt{v}\ul\Psi$ is negligible for $n>1$ because
$\ul{\Psi}$ is slowly varying and we are in the low density, large
scale-factor regime. Then we find that the LQC Hamiltonian
constraint (\ref{qham6}) reduces to a rather simple differential
equation:
\begin{align} \partial_\T^2 \ul{\Psi}(\l_1,\l_2,\l_3;\T) =& \f{8\pi G}{\sqrt v}
\Big[\l_1\f{\partial}{\partial\l_1}\l_2\f{\partial}{\partial\l_2}+\l_1
\f{\partial}{\partial\l_1}\l_3\f{\partial}{\partial\l_3}+\l_2\f{\partial}
{\partial\l_2}\l_1\f{\partial}{\partial\l_1}+\l_2\f{\partial}{\partial\l_2}
\l_3\f{\partial}{\partial\l_3}\nonumber \\& \qquad
+\l_3\f{\partial}
{\partial\l_3}\l_1\f{\partial}{\partial\l_1}+\l_3\f{\partial}{\partial\l_3}
\l_2\f{\partial}{\partial\l_2}\Big]\big(\sqrt{v}\,\ul{\Psi}(\l_1,\l_2,\l_3;\T)
\big). \end{align}
This equation can be further simplified by introducing $\sigma_i =
\log \l_i$ and $\ul{\Phi} = \sqrt{v}\ul\Psi$. The result is:
\be \partial_\T^2 \ul{\Phi}(\sigma_1,\sigma_2,\sigma_3;\T) = 16\pi
G\Big[\f{\partial^2}{\partial\sigma_1\partial\sigma_2}+\f{\partial^2}
{\partial\sigma_1\partial\sigma_3}+\f{\partial^2}{\partial\sigma_2\partial
\sigma_3}\Big] \ul{\Phi} (\sigma_1,\sigma_2,\sigma_3;\T)\, , \ee
where $v$ is now given by $2\exp (\sum\sigma_i)$. This is precisely
the equation we would have obtained if we had started from the
classical Hamiltonian constraint, used the Schr\"odinger
quantization and the ``covariant factor ordering'' of the constraint
as in the \WDW theory. Thus, the LQC Hamiltonian constraint reduces
to the \WDW equation under the assumption that $\ul\Psi$ is slowly
varying in the sense that $(1/\l_i\l_j)^n
\partial_k^n \sqrt{v}\ul\Psi$ can be neglected for $n>1$ relative to
the term for $n=1$. Since $(\l_i\l_j)^2$ is essentially the area
of the $i$-$j$ face of the fiducial cell $\V$ in Planck units,
this should be an excellent approximation well away from the
Planck regime. However, in the Planck regime itself the terms
which are neglected in the LQC dynamics are comparable to the
terms which are kept whence, as in the isotropic case, the \WDW
evolution completely fails to approximate the LQC dynamics.

\section{Discussion}
\label{4}

In this paper we extended the ``improved'' LQC dynamics of Friedmann
space-times \cite{aps3} to obtain a coherent quantum theory of
Bianchi I models. As in the isotropic case, we restricted the matter
source to be a massless scalar field since it serves as a viable
relational time parameter (a la Leibniz) both in the classical and
quantum theories. However, it is rather straightforward to
accommodate additional matter fields in this framework.

To incorporate the Bianchi I model, we had to overcome several
significant obstacles. First, using discrete symmetries we showed
that to specify dynamics it suffices to focus just on the positive
octant. This simplified our task considerably. Second, in section
\ref{s3.2} we introduced a more precise correspondence between LQG
and LQC and used it to fix the parameters $\bar\mu_i$ that determine
the elementary plaquettes, holonomies around which define the
curvature operator $\hat{F}_{ab}{}^k$. This procedure led us to the
expressions $\bar\mu_1^2 = (|p_1|\Delta\,\lp^2)/|p_2 p_3|$, etc.
They reduce to the expression $\bar\mu^2= (\Delta\,\lp^2)/|p|$ of
the isotropic models \cite{aps3,apsv,kv2}. But even there, the
current reasoning has the advantage that it uses only quantum
geometry, avoiding reference to classical areas even in the
intermediate steps. However, because of this rather complicated
dependence of $\bar\mu_i$ on $p_i$, the task of defining operators
$\sin \bar\mu_ic_i$ seems hopelessly difficult at first. Indeed,
this was the key reason why the earlier treatments
\cite{chiou,cv,szulc} took a short cut and simply set $\bar\mu_i^2 =
(\Delta\,\lp^2)/|p_i|$ by appealing to the relation $\bar\mu^2 =
(\Delta\,\lp^2)/|p|$ in the isotropic case. With this choice,
quantization of the Hamiltonian constraint became straightforward
and the final Bianchi I quantum theory resembled three copies of
that of the Friedmann model. However, this result had the physically
unacceptable consequence that significant departures from general
relativity could occur in ``tame'' situations. By a non-trivial
extension of the geometrical reasoning used in the isotropic case,
in section \ref{s3.3} we were able to define the operators $\sin
\bar\mu_i c_i$ for our expressions of $\bar\mu_i$. However, the
structure of the resulting Hamiltonian constraint turned out to be
rather opaque. To simplify its form, in section \ref{s3.4} we
introduced volume as one of the arguments of the wave functions. The
action of the gravitational part of the Hamiltonian constraint then
became transparent: it turned out to be a difference operator where
the multiplicative coefficients in individual terms depend only on
volume and the change in the arguments of the wave functions also
depends only on volume; individual anisotropies do not feature (see
(\ref{qham6}) - (\ref{qham8})). This simplification enabled us to
show that the sector $\H_{\rm reg}^{\g}$ of quantum states which
have no support on classically singular configurations is preserved
by quantum dynamics. In this precise sense the big-bang singularity
is resolved. Furthermore, this quantum dynamics is free from the
physical drawbacks of the older scheme mentioned above.

In section \ref{s4} we explored three consequences of quantum
dynamics in some detail. First, we showed that there is a
projection map $\P:\,\,\H_{\gen} \rightarrow \H_{\res}$ from the
Hilbert space of the more general Bianchi I model to that of the
more restricted Friedmann model which maps the Bianchi I quantum
constraint \emph{exactly} to the Friedmann quantum constraint.
This is possible because, as noted above, it is just the volume
---rather than the anisotropies--- that govern the action of the
Bianchi I quantum constraint. This result is of considerable
interest because, in view of the BKL conjecture, it suggests that
near generic space-like singularities the LQC of Friedmann models
may capture qualitative features of the \emph{full, LQG} dynamics
of the isotropic, homogeneous degree of freedom. In section
\ref{s4.2} we briefly recalled the effective equations of Chiou
and Vandersloot (see Appendix C of \cite{cv}). These equations
provide intuition for the rich structure of quantum bounces in the
Bianchi I model. Their analysis suggests that classical general
relativity is an excellent approximation away from the Planck
regime. However, in the Planck regime quantum geometry effects
rise steeply and forcefully counter the tendency of the classical
equations to drive the matter density, the Ricci scalar and Weyl
invariants to infinity. (In particular, as in the isotropic case,
the matter density is again bounded above by $\rcr \approx 0.41
\rp$.) Thus the quantum geometry effects dilute these quantities
and, once the quantity exits the Planck regime, classical general
relativity again becomes an excellent approximation. In section
\ref{s4.3} we showed that, as in the isotropic case
\cite{aps3,apsv,acs}, there is a precise sense in which LQC
dynamics is well approximated by that of the \WDW theory once
quantum geometry effects become negligible.

The rather complicated dependence of $\bar\mu_i$ on $p_i$ is also
necessary to remove a fundamental conceptual limitation of the
older treatments of the Bianchi I model. Recall that, because we
have homogeneity and the spatial topology is non-compact, we have
to introduce a fiducial cell $\V$ to construct a Lagrangian or a
Hamiltonian framework. Of course, the final physical results must
be independent of this choice. At first this seems like an
innocuous requirement but it turns out to be rather powerful. We
will now recall from \cite{szulc} the argument that this condition
is violated with the simpler choice $\bar\mu_i^2 =
(\Delta\,\lp^2)/|p_i|$ but respected by the more complicated
choice we were led to from LQG.

For definiteness, let us fix a fiducial metric $\q_{ab}$ and
denote by $L_i$ the lengths of the edges of the fiducial cell
$\V$. Suppose we were to use a different cell, $\V^\prime$ whose
edges have lengths $L^\prime_i= \beta_iL_i$ (no summation over
$i$). Since the basic canonical fields $A_a^i$ and $E^a_i$ are
insensitive to the choice of the cell, Eq. (\ref{var}) implies
that the labels $c_i$ and $p_i$ we used to characterize them
change to $c^\prime_1 = \beta_1 c_1$, $p^\prime_1 = \beta_2\beta_3
p_1$, etc. The gravitational part of the classical Hamiltonian
constraint (\ref{ham3}) is just rescaled by an overall factor
$(\beta_1\beta_2\beta_3)^2$ and the inverse symplectic structure
is rescaled by $(\beta_1\beta_2\beta_3)^{-1}$. Hence the
Hamiltonian vector field is rescaled by $(\beta_1\beta_2\beta_3)$,
exactly as it should because the lapse is rescaled by the same
factor. Thus, as one would expect, the classical Hamiltonian flow
is insensitive to the change $\V \rightarrow \V^\prime$. What is
the situation in the quantum theory? Physical states belong to the
kernel of the Hamiltonian constraint operator
$\hat{\mathcal{C}}_H$ whence the two quantum theories will carry
the same physics only if $\hat{\mathcal{C}}_H$ is changed at most
by an overall rescaling. Analysis is a bit more involved than in
the classical case because $\hat{\mathcal{C}}_\g$ involves factors
of $\sin \bar\mu_ic_i$. Now, under $\V \rightarrow \V^\prime$, our
$\bar\mu_i$ transform as $\bar\mu_1 \rightarrow \bar\mu_1^\prime =
\beta_1^{-1}\bar\mu_1$, whence $\bar\mu^\prime_1c^\prime_1 =
\bar\mu_1c_1$, etc, and the Hamiltonian constraint (\ref{qham1})
is rescaled by an overall multiplicative factor
$(\beta_1\beta_2\beta_3)^2$ just as in the classical theory. What
happens if we set $\bar\mu_i^2 = \Delta\,\lp^2/|p_i|$ as in
\cite{chiou,cv,szulc}? Then, we are led to $\bar\mu_1^\prime
c_1^\prime = (\beta_1/\sqrt{\beta_2\beta_3})\, \bar\mu_1c_1$ etc.
Since the constraint (\ref{qham1}) is a sum of terms of the type
$p_1p_2 |p_3| \sin\bar{\mu}_1 c_1\, \sin\bar{\mu}_2c_2$ it has a
rather uncontrolled transformation property and is not simply
rescaled by an overall factor. It is then not surprising that, in
the Planck regime, the dynamical predictions of the resulting
quantum theory (as well as of the effective theory) depend on the
choice of the elementary cell. It is rather remarkable that the
more complicated form of $\bar\mu_i$ that we are led to from LQG
kinematics has exactly the right form to make quantum dynamics
insensitive to the choice of the fiducial cell $\V$. As mentioned
above, it also ensures that the predictions of quantum theory is
free of drawbacks of the earlier treatments \cite{chiou}, such as
the correlation between the bounce and ``directional densities''
which do not have an invariant significance.

From physical considerations, as in the isotropic case, it would be
most interesting to start at a ``late time'' with states that are
sharply peaked at a classical solution in which the three scale
factors assume values for which the curvature is ``tame'' and $\pT$
is very large compared to $\hbar$ in classical units $c$=$G$=1. One
would then evolve these states backward and forward in the
``internal'' time $\T$. As we just discussed, analytical
considerations show that, since the initial wave function is in
$\H_{\rm reg}^{\g}$, it will continue to be in that sub-space; there
is no danger that the expectation values of curvature, anisotropies
or density would diverge. But several important questions remain.
Are there quantum bounces with a pre-big-bang branch again
corresponding to a large, classical universe in the distant past? Is
there is a clear distinction between evolutions of data in which
there are significant initial anisotropies and data which represent
only perturbations on isotropic situations? Even in the second case,
do anisotropies grow (or decay) following predictions of the
classical theory or are there noticeable deviations because of
accumulations of quantum effects over large time periods? Numerical
simulations of the LQC equations are essential to provide confidence
in the general scenario suggested by effective equations and to
supply us with detailed Planck scale physics.

Finally, let us return to full LQG. At the present stage of
development, there appears to be considerable freedom in the
definition of the quantum Hamiltonian constraint in the full theory.
Furthermore, our current understanding of the \emph{physical
implications} of these choices is quite limited. Already in the
isotropic models, the ``improved'' dynamics scheme provided some
useful lessons: it brought out the fact that these choices can be
non-trivially narrowed down by carefully analyzing conceptual issues
(e.g., requiring that the physical results should be independent of
auxiliary structures introduced in the intermediate steps) and by
working out the physical consequences of the theory in detail (to
ensure that the quantum geometry effects are not dominant in the low
energy regime). Rather innocuous choices ---such as those made in
arriving at the older ``$\mu_o$-scheme''--- can lead to unacceptable
consequences on both these fronts \cite{cs}. The Bianchi I analysis
has sharpened these lessons considerably. The fact that the
kinematical interplay between LQG and LQC has a deep impact on the
viability of quantum dynamics is especially revealing. A quantum
analysis of inhomogeneous perturbations around Bianchi I backgrounds
is therefore a promising direction for understanding the physical
implications of the choices that have to be made in the definition
of the Hamiltonian constraint in full LQG. Such an analysis is
likely to narrow down choices and lead us to viable quantization
schemes in LQG that lead to a good semi-classical behavior.

\section*{Acknowledgements:}

We would like to thank Martin Bojowald, Miguel Campiglia,
Alejandro Corichi, Adam Henderson, Frank Hermann, Mercedes
Mart\'in-Benito, Guillermo Mena Marug\'an, Tomasz Pawlowski, Param
Singh, David Sloan, Lukasz Szulc, Manuel Tiglio, and especially
Dah Wei Chiou and Kevin Vandersloot for stimulating discussions.
This work was supported in part by the NSF grants PHY04-56913 and
PHY0854743, Le Fonds qu\'eb\'ecois de la recherche sur la nature
et les technologies, The George A. and Margaret M. Downsbrough
Endowment and the Eberly research funds of Penn State.

\begin{appendix}

\section{Parity Symmetries}%in $\hat{\mathcal{C}}_H$}
\label{a1}

In this appendix we recall and extend results on parity symmetries
obtained in \cite{bd}.

In non-gravitational physics, parity transformations are normally
taken to be discrete diffeomorphisms $x_i \rightarrow -x_i$ in the
physical space which are isometries of the flat 3-metric thereon.
In the phase space formulation of general relativity,  we do not
have a flat metric ---or indeed, any fixed metric. However, if the
dynamical variables have internal indices ---such as the triads
and connections used in LQG--- we can use the fact that the
internal space $I$ is a vector space equipped with a flat metric
$q_{ij}$ to define parity operations on the internal indices.
Associated with any unit \emph{internal} vector $\xi^I$, there is
a parity operator $\Pi_\xi$ which reflects the internal vectors in
the 2-plane orthogonal to $\xi$. This operation induces a natural
action on triads $e^a_i$, the connections $A_a^i$ and the
conjugate momenta $P^a_i =: (1/8\pi G\gamma) E^a_i$ (since they
are internal vectors or co-vectors). It turns out that $e^a_i$ are
proper \emph{internal} co-vectors while $A_a^i$ and $P^a_i$ are
pseudo \emph{internal} vectors and co-vectors, respectively. These
geometrical considerations show that the Barbero-Immirzi parameter
$\gamma$ must change sign under any one of these parity
operations, i.e., if it has the value $|\gamma|$ for say,
positively oriented triads, it should have the value $-|\gamma|$
for negatively oriented triads. Its value on degenerate triads is
ambiguous so on the degenerate sector we cannot unambiguously
recover the triads $e^a_i$ from the momenta $P^a_i$. If one were
to make $\gamma$ a dynamical field \cite{ty,mt}, it follows that
the field should be a \emph{pseudo}-scalar under \emph{internal}
parity transformations; geometrical considerations involving
torsion have led to the same conclusion in \cite{mt}. (For
details, see \cite{aa-dis}).

In the diagonal Bianchi I model, we can restrict ourselves just to
three parity operations $\Pi_i$. Under their action, the canonical
variables $c_i,p_i$ transform as follows:
\be \label{P1} \Pi_1 (c_1,c_2, c_3) = (c_1, -c_2, -c_3), \quad
\quad \Pi_1 (p_1, p_2, p_3) = (-p_1, p_2, p_3)\, , \ee
and the action of $\Pi_2, \Pi_3$ is given by cyclic permutations.
Under any of these maps $\Pi_i$, the Hamiltonian (\ref{ham3}) is
left invariant. This is just as one would expect because $\Pi_i$ are
simply large gauge transformations of the theory under which the
physical metric $q_{ab}$ and the extrinsic curvature $K_{ab}$ do not
change. It is clear from the action (\ref{P1}) that if one knows the
dynamical trajectories on the octant $p_i\ge 0$ of the phase space,
then dynamical trajectories on any other octant can be obtained just
by applying a suitable (combination of) $\Pi_i$. Therefore, in the
classical theory one can restrict one's attention just to the
positive octant.

Let us now turn to the quantum theory. We now have three operators
$\hat\Pi_i$. Their action on states is given by
\be \hat\Pi_1 \Psi(\l_1,\l_2,\l_3) = \Psi(-\l_1,\l_2\l_3)\, , \ee
etc. What is the induced action on operators? Since
\begin{align} \label{P2}\hat{\Pi}_1\l_1\hat{\Pi}_1\Psi(\l_1,\l_2,\l_3)
&= \hat{\Pi}_1 \Big(\l_1\,\Psi(-\l_1,\l_2,\l_3)\Big) \nonumber \\
&= -\l_1\Psi(\l_1,\l_2,\l_3), \end{align}
we have
\be \label{P3} \hat{\Pi}_1\l_1\hat{\Pi}_1  = -\l_1. \ee
The Hamiltonian constraint operator is given by Eqs. (\ref{qham2})-
(\ref{qham3}). To calculate its transformation property under parity
maps, in addition to (\ref{P3}), we also need the transformation
property of operators $\sin \bar\mu_ic_i$. An inspection of Eq.
(\ref{qham3}) shows that, in view of the Bianchi I symmetries,  it
is sufficient to calculate $\hat{\Pi}_i \sin \bar\mu_1c_1
\hat{\Pi}_i$.  We have:
\begin{align} \hat{\Pi}_1\sin\bar\mu_1c_1\hat{\Pi}_1\Psi(\l_1,\l_2,\l_3)
&= \f{1}{2i}\, \hat\Pi_1\,\Big[\Psi(-\l_1-\f{\sgn(-\l_1)}{\l_2\l_3},\l_2,\l_3)-
\Psi(-\l_1+\f{\sgn(-\l_1)} {\l_2\l_3},\l_2,\l_3)\Big] \nonumber \\
&= \f{1}{2i}\Big[\Psi(\l_1-
\f{\sgn(\l_1)}{\l_2\l_3},\l_2,\l_3)-\Psi(\l_1+\f{\sgn(\l_1)}{\l_2\l_3},\l_2,
\l_3)\Big] \nonumber \\ &= \sin\bar\mu_1c_1\Psi(\l_1,\l_2,\l_3), \end{align}
whence
\be \hat{\Pi}_1 \sin\bar\mu_1c_1\hat\Pi_1 = \sin\bar\mu_1c_1. \ee
An identical calculation shows that
\begin{align} \hat{\Pi}_2\sin\bar\mu_1c_1\hat{\Pi}_2\,\Psi(\l_1,\l_2,\l_3)
&= \f{1}{2i}\, \hat\Pi_2\, \Big[ \Psi(\l_1-\f{\sgn(\l_1)}{(-\l_2)\l_3},
-\l_2,\l_3)-\Psi(\l_1+\f{\sgn(\l_1)}{(-\l_2)\l_3},-\l_2,\l_3)\Big] \nonumber\\
&= \f{1}{2i}\Big[\Psi(\l_1+ \f{\sgn(\l_1)}{\l_2\l_3},\l_2,\l_3)-
\Psi(\l_1-\f{\sgn(\l_1)}{\l_2\l_3},\l_2,\l_3)\Big] \nonumber \\
&= -\sin\bar\mu_1c_1\Psi(\l_1,\l_2,\l_3)\, , \end{align}
and similarly for $\hat\Pi_3$. Therefore, we have:
\be \hat{\Pi}_2 \sin\bar\mu_1c_1\hat\Pi_2 = - \sin\bar\mu_1c_1,
\quad {\rm and} \quad \hat{\Pi}_3 \sin\bar\mu_1c_1\hat{\Pi}_3 =
-\sin\bar\mu_1c_1.\ee
These transformation properties of $\sin\bar\mu_1c_1$ under
$\hat\Pi_i$ simply mirror the transformation properties of $c_1$
under the three parity operations $\Pi_i$ in the classical theory.
(Note that, because of the absolute value signs in the expressions
(\ref{mubar}), $\bar\mu_i$ do not change under any of the parity
maps.)

From Eqs. (\ref{qham2})- (\ref{qham3}) it now immediately follows
that the gravitational part of the Hamiltonian constraint is left
invariant under $\hat\Pi_i$. Since $\hat{p}_{(\T)}^2$ is manifestly
invariant, we have:
\be \hat{\Pi}_i\,\, \hat{\mathcal{C}}_H \,\,\hat{\Pi}_i =
\hat{\mathcal{C}}_H. \ee
just as in the classical theory. Because of this invariance
property, given any state $\Psi \in \Hkg$, the restriction to the
positive octant of its image under $\hat{\mathcal{C}}_\g$
determines its image everywhere on $\Hkg$. As we saw in section
\ref{s3.4}, this property simplifies the task of finding the
explicit action of the Hamiltonian constraint considerably.

\end{appendix}

\end{document}